\documentstyle[amssymb,preprint,aps,prb]{revtex}

\tighten
\ifpreprintsty 
\makeatletter
\def\@CITEX[#1]#2{\if@filesw\immediate\write\@auxout{\string\citation{#2}}\fi
\leavevmode [\@cite{\@collapse{#2}}{#1}]}
\makeatother
\fi

\begin{document}
\draft

\title{\centerline{Multi-band Gutzwiller wave functions for general
on-site interactions}}

\author{J.~B\"unemann and W.~Weber}
\address{Inst.~f.~Physik, Universit\"at Dortmund, D-44221 Dortmund, Germany}
\author{F.~Gebhard}
\address{Inst.\ Laue--Langevin, B.~P.\ 156x, F-38042 Grenoble Cedex 9, France}
\date{\bf Accepted for publication in PRB~{\bf 57} (March 15, 1998)}  
\maketitle

\begin{abstract}%
We introduce Gutzwiller wave functions for 
multi-band models with general on-site Coulomb interactions.
As these wave functions employ correlators for the exact atomic eigenstates
they are exact both in the non-interacting and in the atomic limit. 
We evaluate them in infinite lattice dimensions 
for all interaction strengths without
any restrictions on the structure of the Hamiltonian or the symmetry of the
ground state. The results for the ground-state energy allow us to derive an
effective one-electron Hamiltonian for Landau quasi-particles, applicable
for finite temperatures and frequencies within the Fermi-liquid regime.

As applications for a two-band model 
we study the Brinkman--Rice metal-to-insulator
transition at half band-filling, and the transition to itinerant
ferromagnetism for two specific fillings, at and close to a peak
in the density of states of the non-interacting system.
Our new results significantly differ from those for
earlier Gutzwiller wave functions
where only density-type interactions were included.
When the correct spin symmetries for the two-electron states are
taken into account, the importance of the Hund's-rule exchange
interaction is even more pronounced and leads to paramagnetic
metallic ground states with large local magnetic moments.
Ferromagnetism requires fairly large interaction strengths,
and the resulting ferromagnetic state is a strongly correlated metal.
\end{abstract}


\section{Introduction}
\label{Intro}

In transition metals and their compounds the electrons of the open 
$d$~shells participate in the itineracy of the valence
electrons. At the same time their atomic correlations,
as described, e.g., by Hund's first and second rule, remain important.
The competition between the electrons'
itinerant and local features results in many interesting phenomena:
magnetism is most prominent, but also metal-to-insulator transitions, 
high-to-low-spin changes, orbital ordering etc.\ occur in these materials.

For insulating compounds it is commonly accepted that their magnetic behavior
can be described in terms of the spins of {\em localized\/}
electrons which are coupled by superexchange via the ligands.
However, there are conflicting views on the magnetism of the metallic state.
One line of reasoning which dates back to van Vleck and others
assumes rather small charge fluctuations around the average (atomic)
$d^n$~configuration, and the localized spins remain a useful concept
also in the metallic state (minimum polarity model~\cite{vanVleck}).
The other school, starting from the Hartree--Fock--Stoner theory,
treats magnetism within a single-particle band theory, i.e.,
in a completely itinerant limit.
In particular, spin-density functional theory quite successfully describes
the ferromagnetism of the iron group metals, 
not only concerning magnetic moments but also such 
details as the shapes of complicated multi-sheet 
Fermi surfaces.~\cite{Moruzzi} 
In the spirit of a free-electron theory the spin-density functional
theory generally assumes a local exchange-correlation potential
which is a function of the local charge and spin densities.
The success of this effective single-particle theory is quite surprising
since, in the atomic limit, it cannot reproduce the open-shell
electronic structure.

In his seminal work Gutzwiller~\cite{Gutzwiller1963} proposed a variational
approach to the problem of itinerant ferromagnetism in the 
Hubbard model.~\cite{Hubbard} 
In his many-body trial state atomic configurations with large deviations from 
the average occupancy could be reduced with respect to
a Hartree--Fock reference state, depending on the value of
the variational parameter. Therefore, his approach incorporates both
the itinerant and localized aspects of itinerant magnetism.
Gutzwiller introduced an approximate evaluation of his many-body 
wave function, the so-called Gutzwiller approximation, 
and concluded from his one-band results that itinerant
ferromagnetism requires large interaction strengths. Later, Brinkman and
Rice realized~\cite{BrinkmanRice} that the Gutzwiller approximation
contained a transition from a paramagnetic metal to a paramagnetic insulator
in which all electrons are localized. The Brinkman--Rice insulator provides
an instructive example for the more general class of Mott--Hubbard
insulators.~\cite{Mottbuch,BUCH}

% NEW November 26, 1997
During the last decade new analytical techniques were developed which allow 
for an exact evaluation of the single-band Gutzwiller wave function in one 
dimension,~\cite{MVGV} and in the limit of infinite 
dimensions.~\cite{MVPRLdinfty,METZNER} For the latter case
the results of the Gutzwiller approximation were found to become exact.
Furthermore, the results of the
Kotliar--Ruckenstein Slave-Boson mean-field theory~\cite{KRPRL} were re-derived
for the paramagnetic and the antiferromagnetic 
case.~\cite{MVPRLdinfty,METZNER}
In this work we extend the single-band formalism of
Refs.~\onlinecite{GebhardPRB1,GebhardPRB2}, 
which provided exact results in infinite dimensions for the whole
class of Gutzwiller wave functions. For the one-band case the Gutzwiller
variational approach in infinite dimensions 
and Slave-Boson theories~\cite{KRPRL,Piers} on mean-field level
were shown to be completely equivalent;
see Ref.~\onlinecite[Chap.~3]{BUCH} for a recent review.
% NEW November 26, 1997

The case of multi-band systems poses a more complicated problem. 
Each lattice site represents an atom with an incompletely filled shell.
Consequently, the atomic Hamiltonian $\hat{H}_{\text{at}}$ 
should include the relevant properties of the electronic structure of 
isolated atoms or ions.
This means that it should comply with Hund's first rule,
it should reproduce the essential features of low-lying multiplet excitations,
and it has to incorporate the symmetry of the ligand field.
The commonly used form for the atomic part~$\hat{H}_{\text{at}}$
of the multi-band Hubbard Hamiltonian includes:
(i)~an orbital-diagonal density-density interaction of strength~$U$ 
as in the single-band case,
and an orbital-nondiagonal density-density interaction
of strength~$U'$;
(ii)~two-particle spin-exchange terms of strength~$J$ such that
the ground state of $\hat{H}_{\text{at}}$ fulfills Hund's first rule,
i.e., it exhibits maximum spin for $J>0$.
Frequently, the spin-exchange terms are taken into account only partially.
If the (orbital) spin-flip terms are neglected 
we are left with~$\hat{H}_{\text{at}}^{\text{dens}}$
which contains only density-density interactions on the atomic sites.

% NEW November 26, 1997
It should be noted that a while ago atomic
Hamiltonians similar to~$\hat{H}_{\text{at}}$ were studied using the 
``local ansatz'',~\cite{FULDEBUCH,STOLLHOFF} a scheme in the spirit of the
Gutzwiller method.
However, the results presented were limited to small interaction strengths.
Recently, the Gutzwiller method was generalized to treat
multi-band Hubbard models with local density-type interactions
$\hat{H}_{\text{at}}^{\text{dens}}$ of arbitrary
strengths.~\cite{BWa,BWb,Okabe,BGW}
% NEW November 26, 1997
The essential idea was to evaluate correlators for atomic multi-electron
configurations made up of spin-orbital product
states (`Slater determinants'). 
This was possible since $\hat{H}_{\text{at}}^{\text{dens}}$ 
is diagonal in these configurations. In a first step the Gutzwiller 
approximation was used,~\cite{BWa,BWb,Okabe} later it was shown that 
these results become exact in the limit of infinite dimensions.~\cite{BGW} 
In Ref.~\onlinecite{BGW} we compared our results 
with those of previous generalizations of the Gutzwiller approximation 
to the case of degenerate bands.

However, the frequently used treatment of $\hat{H}_{\text{at}}$ as discussed 
above still {\em violates\/} the atomic symmetry; for an example, see below.
The reason is the incomplete form of the exchange interaction.
To establish the correct symmetry it is necessary to include
{\em all\/} exchange terms which result from a (spin-conserving)
two-particle interaction, i.e., we will have to consider 
the contributions from up to four different spin-orbits. 
Then, the proper $\frak{n}$-electron
atomic eigenstates are certain linear combinations of the respective
$\frak{n}$-electron spin-orbit product configurations.
As a consequence, the optimum way to generalize the
Gutzwiller wave function to multi-band systems is 
the use of correlators for the atomic $\frak{n}$-electron eigenstates
instead of the pure spin-orbit product states.

In this paper we introduce and evaluate such variational wave functions 
with atomic correlations. Our formulation allows for arbitrary orbital bases, 
including more than one orbital type per representation, i.e.,
more than one type of $s$, $p$, or $d$ orbitals. 
It also allows for an arbitrary number of atomic sites in the unit cell. 
In the limit of infinite dimension, exact results for the
ground-state energy are given in terms of an effective single-particle 
Hamiltonian which defines the band structure of correlated electrons.
Thus, our theory naturally extends to finite temperatures 
(Fermi-liquid regime). As an example we apply our theory to a two-band model
and show that the correct treatment of the atomic correlations yields a
variety of results which quantitatively and, in some cases, even
qualitatively differ from those using pure density correlations.

Our paper is structured as follows. In Sect.~\ref{Hamiltonians} we introduce
the multi-band Hubbard Hamiltonian with purely on-site interactions. The
spectrum of the general atomic Hamiltonian is supposed to be known. Then, in
Sect.~\ref{wave_functions}, we specify the class of Gutzwiller wave
functions with atomic correlations and give the exact results for the
ground-state energy in infinite dimensions. For the case of pure density
correlations we recover our previous expressions.~\cite{BWa,BWb,BGW} In
Sect.~\ref{twodegbands}, we discuss the example of two partly filled $e_g$
bands in more detail. We study the Brinkman--Rice metal-insulator transition
at half band-filling, and itinerant ferromagnetism for two generic
band-fillings. A summary and conclusions close our presentation. Technical
details are deferred to the appendix.

\section{Hamilton operator}
\label{Hamiltonians}

\subsection{Multi-band Hubbard model}

\label{itinerantproblem}

Our multi-band Hubbard model~\cite{Hubbard} is defined by the Hamiltonian 
\begin{equation}
\hat{H}=\sum_{i,j;\bbox{\sigma},\bbox{\sigma'}}
t_{i,j}^{\bbox{\sigma},\bbox{\sigma'}}
\hat{c}_{i;\bbox{\sigma}}^{+}
\hat{c}_{j;\bbox{\sigma'}}^{\vphantom{+}}
+ \sum_i\hat{H}_{i;\text{at}}\equiv \hat{H}_1+\hat{H}_{\text{at}}\;.  \label{1}
\end{equation}
Here, $\hat{c}_{i;\bbox{\sigma}}^{+}$ creates an electron with combined
spin-orbit index~$\bbox{\sigma}=1,\ldots ,2N$ ($N=5$ for 3$d$~electrons) at
the lattice site~$i$ of a solid. 
% NEW November 26, 1997
We do not yet specify a periodic lattice, i.e.,
the sites $i$ may also represent ligand atoms.
Therefore, the number of orbitals~$N$ also depends on the site, $N\equiv N_i$.
% NEW November 26, 1997
To keep the notation transparent we will drop
this additional index in the following, and we will use the notion
``orbital'' for spin-orbit states. 

The first term in~(\ref{1}), $\hat{H}_1$, represents an appropriate
single-particle tight-binding Hamiltonian. 
Crystal field terms are included in the orbital energies 
$t_{i,i}^{\bbox{\sigma},\bbox{\sigma}}\equiv
\epsilon_{i;\bbox{\sigma}}$. We may also allow non-diagonal crystal field
terms $t_{i,i}^{\bbox{\sigma},\bbox{\sigma'}}$ 
for $\bbox{\sigma}\neq \bbox{\sigma'}$ in case of a sufficiently 
large orbital basis (or a sufficiently low atomic-site symmetry). 
In this paper we do not include spin-orbit coupling and we may therefore
assume that the terms $t_{i,j}^{\bbox{\sigma},\bbox{\sigma'}}$ 
are spin-independent quantities which only depend
on the spatial part of the underlying spin-orbit wave functions.

In our model~(\ref{1}) we assume that the electrons interact only locally. 
Separating the density-density interactions we may write the atomic Hamiltonian
as two terms,
\begin{equation}
\hat{H}_{i;\text{at}}=\sum_{\bbox{\sigma},\bbox{\sigma'}\,
 (\bbox{\sigma}\neq\bbox{\sigma'})}
{\cal U}_i^{\bbox{\sigma},\bbox{\sigma'}}
\hat{n}_{i;\bbox{\sigma}}\hat{n}_{i;\bbox{\sigma'}}
+\sum_{(\bbox{\sigma_1}<\bbox{\sigma_2})\neq (\bbox{\sigma_3}>\bbox{\sigma_4})}
{\cal J}_i^{\bbox{\sigma_1},\bbox{\sigma_2};\bbox{\sigma_3},\bbox{\sigma_4}}
\hat{c}_{i;\bbox{\sigma_1}}^{+}\hat{c}_{i;\bbox{\sigma_2}}^{+}
\hat{c}_{i;\bbox{\sigma_3}}^{\vphantom{+}}
\hat{c}_{i;\bbox{\sigma_4}}^{\vphantom{+}}\;.  \label{fullHat}
\end{equation}
The exchange-type second term transfers two electrons from the
orbitals $\bbox{\sigma_3}>\bbox{\sigma_4}$ into the 
orbitals $\bbox{\sigma_1}<\bbox{\sigma_2}$.
The quantities ${\cal U}_i$ and ${\cal J}_i$ represent all
possible two-particle (Coulomb) interactions compatible with the symmetry at
site $i$. 

\subsection{The atomic problem}
\label{atomicDEF}

In our variational wave functions we will deal with operators which project
onto atomic ${\frak n}$-electron eigenstates $|\Gamma_i\rangle $ with
arbitrary number $0\leq {\frak n}\leq 2N$. Altogether we will have $2^{2N}$
eigenstates. Each of these ${\frak n}$-electron eigenstates has the proper
symmetry, i.e., they can be classified according to irreducible
representations of the group defined by the symmetry of site $i$. Part of
this classification is according to the total spin quantum number
as $\hat{H}_{i;\text{at}}$ commutes with $(\vec{S}_i)^2$. The problem of
classification is treated in detail by many 
authors;~\cite{Griffith,Ballhausen} here we refer to the
book of Sugano et al.~\cite{Sugano}

We now suppress the site index, and introduce the following notation for all
possible $2^{2N}$ multi-orbital configurations $I$ and the corresponding
multi-electron configuration states $|I\rangle $.

\begin{enumerate}
\item  A configuration~$I$ is characterized by the electron occupation of
the orbitals, 
\begin{equation}
I\in \left\{ \emptyset ;(1),\ldots ,(2N);(1,2),\ldots ,(2,3),\ldots
(2N-1,2N);\ldots ;(1,\ldots ,2N)\right\} \;.  \label{2}
\end{equation}
Here the symbol $\emptyset $ in~(\ref{2}) means that the site is empty.
Then, there follow all $2N$ one-electron configurations, all $N(2N-1)$
two-electron configurations --the sequence of numbers in round brackets 
$(\bbox{\sigma_{1}},\bbox{\sigma_{2}},\ldots)$ is irrelevant-- and so on up to
the $2N$~electron configuration $(1,\ldots ,2N)$. 
For example, $(\bbox{\sigma_1},\bbox{\sigma_2})$ specifies one of
the 45~possible $[\text{Ar}]\, 3d^2$ configurations of a 
Ti$^{++}$ ion in the frozen-core approximation.

In general, we interpret
the indices~$I$ in~(\ref{2}) as sets in the usual sense. For example, in the
atomic configuration $I\backslash I'$ only those orbitals in~$I$
are occupied which are not in~$I'$. The complement of~$I$ is 
$\overline{I}=(1,2,\ldots ,2N)\backslash I$, i.e., in the atomic
configuration $\overline{I}$ all orbitals but those in~$I$ are occupied.
\item  $|I|\equiv {\frak n}$, i.e., the absolute value $|I|$ of a
configuration indicates the number ${\frak n}$ of a multi-electron state,
\begin{equation}
|\emptyset |=0;|(\bbox{\sigma_1})|=1;
|(\bbox{\sigma_1},\bbox{\sigma_2})|=2;
\ldots ;|(1,\ldots ,2N)|=2N\;.  \label{3}
\end{equation}
\item  A multi-electron configuration state (`Slater determinant')
is constructed as 
\begin{equation}
|I\rangle =|\bbox{\sigma_1},\bbox{\sigma_2},\ldots ,\bbox{\sigma_{|I|}}\rangle
=\prod_{n=1}^{|I|}\hat{c}_{\bbox{\sigma_n}}^{+}|\text{vacuum}\rangle  
\quad (\bbox{\sigma_n}\in I) \; . 
\end{equation}
The sequence of electron creation operators in~$|I\rangle $ is in ascending
order, i.e., $\bbox{\sigma_i}<\bbox{\sigma_j}$ for $i<j$. When we add an
electron to the configuration eigenstate~$|I\rangle $ with the help of the
electron creation operator we obtain the configuration 
eigenstate $|I\cup \bbox{\sigma}\rangle $ up to the fermionic sign function 
\begin{mathletters}
\label{deffsign}
\begin{equation}
\text{fsgn}(\bbox{\sigma},I)=\langle I\cup \bbox{\sigma}
|\hat{c}_{\bbox{\sigma}}^{+}|I\rangle \;.
\end{equation}
It gives a minus (plus) sign if it takes an odd (even) number of
anticommutations to shift the operator $\hat{c}_{\bbox{\sigma}}^{+}$ to its
proper place in the sequence of electron creation operators 
in $|I\cup \bbox{\sigma}\rangle $. In general, we define for 
$I\cap I^{\prime}=\emptyset$ 
\begin{equation}
\text{fsgn}(I',I)=\langle I\cup I'|
\prod_{{{n=1\vphantom{n_i}} \atop {(\bbox{\sigma_n}\in I')}}}^{|I'|}
\hat{c}_{\bbox{\sigma_n}}^{+}|I\rangle \;.
\end{equation}\end{mathletters}% HERE!
\item  The operator which projects onto a specific configuration~$I$ is
given by 
\begin{mathletters}
\label{g1}
\begin{equation}
\hat{m}_I\equiv \hat{m}_{I,I}=|I\rangle \langle I|=\prod_{\bbox{\sigma}\in I}
\hat{n}_{\bbox{\sigma}}
\prod_{\bbox{\sigma}\in \overline{I}}(1-\hat{n}_{\bbox{\sigma}})\;,  \label{4b}
\end{equation}
where the operators $\hat{m}_I$ fulfill the local completeness relation 
\begin{equation}
\sum_I\hat{m}_I=\openone \;.
\end{equation}\end{mathletters}% HERE!
At this point we also define the operators 
\begin{equation}
\hat{n}_{\emptyset} =\openone\quad ;\quad \hat{n}_I=\prod_{\bbox{\sigma}\in I}
\hat{n}_{\bbox{\sigma}}\quad \text{for $|I|\geq 1$}\;,  \label{4}
\end{equation}
which measure the ``gross'' occupancy of the atom. The gross occupancy
operator $\hat{n}_I$ gives a non-zero result when applied to~$|I'\rangle$
only if~$I$ contains electrons in the same orbitals as~$I'$.
However, $I$ and $I'$ need not be identical because $I'$
could contain additional electrons in further orbitals, i.e., only 
$I\subseteq I'$ is required. Each gross (net) operator can be
written as a sum of net (gross) operators 
\begin{mathletters}
\label{g2b}
\begin{eqnarray}
\hat{n}_I&=&\sum_{I'\supseteq I}\hat{m}_{I'}\;,  \label{4c}
\\
\hat{m}_I&=&\sum_{I'\supseteq I}\left( -1\right) ^{|I^{\prime}
\backslash I|}\hat{n}_{I'}\;.  \label{4d}
\end{eqnarray}\end{mathletters}% HERE!
For practical calculations the net operators~$\hat{m}_I$ are more useful
than the gross operators~$\hat{n}_I$ because the former are projection
operators onto a given configuration~$I$, i.e., 
$\hat{m}_I\,\hat{m}_{I'}=\delta_{I,I'}\hat{m}_I$, 
as can be seen from~(\ref{4b}).
\item  For $I\neq I'$ we denote $J=I\cap I'$, 
$I=J\cup I_1$, and $I'=J\cup I_2$ 
with $I_1\cap I_2=\emptyset $. We want to
describe the transfer of $|I_1|=|I_2|$ electrons from the orbitals $I_2$ to
the orbitals $I_1$ whereas the contents of the other $|J|$~orbitals remains
unchanged. The gross operator for the transfer of electrons from $I_2$ to 
$I_1$ is given by 
\begin{equation}
\hat{n}_{I_1,I_2}=\Bigl(\prod_{
{n=1} \atop {(\bbox{\sigma_n}\in I_1)}
}^{
|I_1|}
\hat{c}_{\bbox{\sigma_n}}^{+}
\Bigr)
\Bigl(
\prod_{
{n=1} \atop {(\bbox{\sigma_n}\in I_2)}
}^{
|I_2|}
\hat{c}_{\bbox{\sigma_{|I_2|-n}}}^{\vphantom{+}}\Bigr)
\;.
\end{equation}
With the help of the fermionic sign function~(\ref{deffsign}) the net
operator for this process can be cast into the form 
\begin{equation}
\hat{m}_{I,I'}=|I\rangle \langle I'|=\text{fsgn}(J,I_1)\text{fsgn}(J,I_2)
\biggl[\prod_{\bbox{\sigma}\in J}\hat{n}_{\bbox{\sigma}}
\prod_{\bbox{\sigma}\in \overline{J}\backslash (I_1\cup I_2)}
(1-\hat{n}_{\bbox{\sigma}})\biggr] \hat{n}_{I_1,I_2}\;.
\end{equation}
All these operators are also defined for $|I_1|\neq |I_2|$. Note the useful
relation 
\begin{equation}
\hat{m}_{I_1,I_2}\hat{m}_{I_3,I_4}=\delta_{I_2,I_3}\hat{m}_{I_1,I_4}\;,
\label{projectoridentity}
\end{equation}
which is easily proven with the help of the Dirac representation of the
operators~$\hat{m}_{I,I'}$.
\end{enumerate}
The configuration eigenstates~$|I\rangle $ form a basis of the atomic Hilbert
space. The atomic Hamiltonian~(\ref{fullHat}) is Hermitian and only states
with the same number of electrons are mixed. For the Hamilton matrix 
$(\tensor{H}_{\text{at}})_{I,I'}=\langle I|
\hat{H}_{\text{at}}|I'\rangle $ 
we can find a unitary matrix~$\tensor{T}$ such that 
\begin{mathletters}
\label{g2strich}
\begin{equation}
(\tensor{T})^{+}\tensor{H}_{\text{at}}\tensor{T}=\text{diag}(E_{\Gamma})\;.
\end{equation}
The atomic eigenstates~$|\Gamma \rangle $ obey 
\begin{eqnarray}
|\Gamma \rangle &=&\sum_IT_{I,\Gamma }|I\rangle \;, \\[3pt]
\hat{H}_{\text{at}}|\Gamma \rangle &=&E_{\Gamma} |\Gamma \rangle
\end{eqnarray}
with 
\begin{equation}
\sum_{\Gamma} T_{I,\Gamma }^{\vphantom{+}} T_{\Gamma ,I'}^{+}=\delta_{I,I'}
\quad ,\quad T_{\Gamma,I }^{+}\equiv T_{I,\Gamma}^{*}\;.
\end{equation}\end{mathletters}% HERE!
Since only configuration eigenstates with the same number of electrons mix,
the matrix~$\tensor{T}$ is block-diagonal with $|\Gamma|=|I|$ for each block.

The atomic Hamiltonian can be written as 
\begin{mathletters}
\label{g2}
\begin{equation}
\hat{H}_{\text{at}} = \sum_{\Gamma} E_{\Gamma} \hat{m}_{\Gamma} \; ,
\end{equation}
where the projection operators 
\begin{equation}
\hat{m}_{\Gamma} = |\Gamma \rangle\langle \Gamma | = \sum_{I,I^{\prime}}
T_{I,\Gamma} \hat{m}_{I,I^{\prime}} T_{\Gamma,I^{\prime}}^+  \label{mgamma}
\end{equation}
fulfill the local completeness relation 
\begin{equation}
\sum_{\Gamma} \hat{m}_{\Gamma} = \openone \; .  \label{localcompelete}
\end{equation}\end{mathletters}% HERE!
For $\Gamma=I=\emptyset$ we set $T_{\emptyset,\emptyset}=1$. For 
$|\Gamma|=|I|= 1$ the atomic Hamiltonian does not fix $T_{I,\Gamma}$, and we
may choose them to facilitate the evaluation of expectation values for our
variational wave functions.

\section{Gutzwiller-correlated wave functions}
\label{wave_functions}

\subsection{Gutzwiller wave functions with atomic correlations}

Gutzwiller-correlated wave functions are Jastrow-type wave functions, i.e.,
they are written as the many-particle correlator~$\hat{P}_{\text{G}}$ acting
on a normalized single-particle product state~$|\Phi_0\rangle $, 
\begin{equation}
|\Psi_{\text{G}}\rangle =\hat{P}_{\text{G}}|\Phi_0\rangle \;.
\end{equation}
Expectation values with the single-particle state~$|\Phi_0\rangle $ are
denoted by 
\begin{equation}
O^0\equiv \langle \hat{O}\rangle_0=\langle \Phi_0|\hat{O}|\Phi_0\rangle
\;.
\end{equation}
In general, these expectation values can be calculated easily with the help
of Wick's theorem.~\cite{FetterWalecka} In the following we will assume that
local Fock terms are absent in~$|\Phi_0\rangle $, i.e., 
\begin{equation}
\langle \Phi_0|
\hat{c}_{i;\bbox{\sigma}}^{+}\hat{c}_{i;\bbox{\sigma'}}^{\vphantom{+}}
|\Phi_0\rangle 
=\delta_{\bbox{\sigma},\bbox{\sigma'}}
\langle\Phi_0|
\hat{c}_{i;\bbox{\sigma}}^{+}\hat{c}_{i;\bbox{\sigma}}^{\vphantom{+}}
|\Phi_0\rangle =\delta_{\bbox{\sigma},\bbox{\sigma'}}n_{i;\bbox{\sigma}}^0
\;.  \label{noFOCKcops}
\end{equation}
This is the case when our orbital basis is sufficiently restricted, i.e.,
when we use only one set of orbitals for each irreducible representation of
the group of the site which we consider in $\hat{H}_1$ of eq.~(\ref{1}). For
cubic symmetry this means that we only consider one type of $s$ and/or $p$
and/or $d$ orbitals. In cases of lower symmetry further
restrictions are possible; for example, in tetragonal site symmetry
$s$-type and $d(3z^2-r^2)$ orbitals may mix. 
For the Hamiltonian~(\ref{1}) we thus choose a basis
where the orbitals are not mixed locally, i.e.,
$t_{i,i}^{\bbox{\sigma},\bbox{\sigma'}}= 0$ for 
$\bbox{\sigma}\neq\bbox{\sigma'}$. In the appendix we treat the general case
without these restrictions.

The one-particle state~$|\Phi_0\rangle $ is usually chosen as the ground
state of an effective one-particle Hamiltonian $\hat{H}_1^{\text{eff}}$.
Apart from the simplest cases $\hat{H}_1^{\text{eff}}$ is not identical with 
$\hat{H}_1$; in general $\hat{H}_1^{\text{eff}}$ has a lower symmetry than 
$\hat{H}_1$. In these cases the restriction (\ref{noFOCKcops}) may also fail.

The one-particle wave function contains many configurations which are
energetically unfavorable with respect to the interacting part of the
Hamiltonian. Hence, the correlator~$\hat{P}_{\text{G}}$ is chosen to 
suppress the weight of these configurations to minimize the total 
energy in~(\ref{1}). In the limit of strong correlations the Gutzwiller 
correlator~$\hat{P}_{\text{G}}$ 
should project onto atomic eigenstates. Therefore, the proper multi-band
Gutzwiller wave function with atomic correlations reads 
\begin{eqnarray}
|\Psi_{\text{G}}\rangle &=&\hat{P}_{\text{G}}|\Phi_0\rangle 
=\prod_i\hat{P}_{i;\text{G}}|\Phi_0\rangle \;,  \nonumber \\[6pt]
\hat{P}_{i;\text{G}} 
&=&
\prod_{\Gamma} \lambda_{i;\Gamma }^{\hat{m}_{i;\Gamma}}
=\prod_{\Gamma} \left[ 1+\left( \lambda_{i;\Gamma }-1\right) \hat{m}_{i;\Gamma}
\right] =1+\sum_{\Gamma} 
\left( \lambda_{i;\Gamma}-1\right) \hat{m}_{i;\Gamma }
\;.  \label{GutzcorrdegbandsHund}
\end{eqnarray}
The $2^{2N}$ variational parameters~$\lambda_{i;\Gamma }$ per site are
real, positive numbers. For $\lambda_{i;\Gamma_0}=0$ and all other 
$\lambda_{i;\Gamma }\neq 0$ the atomic configuration~$|\Gamma_0\rangle $ at
site~$i$ is removed from~$|\Phi_0\rangle $. 

Expectation values in Gutzwiller-correlated wave 
functions~$|\Psi_{\text{G}}\rangle $ will be denoted as 
\begin{equation}
O\equiv \langle \hat{O}\rangle =\frac{\langle \Psi_{\text{G}}|\hat{O}|
\Psi_{\text{G}}\rangle }{\langle \Psi_{\text{G}}|\Psi_{\text{G}}\rangle }\;.
\end{equation}
We will frequently use the expectation values for the atomic eigenstates, 
$m_{i;\Gamma }=\langle \hat{m}_{i;\Gamma }\rangle$, and for the gross and
net occupancy operators, $n_{i;I}=\langle \hat{n}_{i,I}\rangle $ and 
$m_{i;I}=\langle \hat{m}_{i;I}\rangle $.

\subsection{Exact results in infinite dimensions}

\label{Exresultsdinftysec}

Even in the one-band case the evaluation of Gutzwiller-correlated wave
functions is a difficult many-particle problem; 
see Ref.~\onlinecite[Chap.~3.4]{BUCH} for a review. 
It can be solved completely in
the limit of infinite dimensions without further approximations. For 
$d\to\infty$ the electron transfer matrix elements between two sites~$i$ 
and~$j$ at a distance $|i-j|=\sum_{l=1}^d |i_l-j_l|$ on a (hyper-)cubic lattice
have to be scaled as~\cite{MVPRLdinfty} 
\begin{equation}
t_{i,j} = \overline{t}_{i,j} \left(\frac{1}{\sqrt{2d}}\right)^{|i-j|} \; ,
\end{equation}
where $\overline{t}_{i,j}$ is independent of the dimension. In this way the
kinetic and the potential energy compete with each other for all~$d$. The
bandwidth of the electrons stays finite, and we may even use the 
proper $d$-dimensional 
density of states~${\cal D}_{\bbox{\sigma},0}(\epsilon)$ 
for our calculations. The essential simplification in the limit of infinite
dimensions lies in the fact that only local properties of the wave function
are needed for the calculation of single-particle properties; 
see Refs.~\onlinecite[Chap.~5]{BUCH}, \onlinecite{VollhardtWS}, 
and~\onlinecite{MEGAREVMOD} for recent reviews on the limit of 
infinite dimensions.

The class of Gutzwiller-correlated wave functions as specified in 
eq.~(\ref{GutzcorrdegbandsHund}) can also be evaluated exactly in the limit of
infinite dimensions. We defer technical details of the calculations to the
appendix, and merely quote the main result of our work at this point. In the
limit of infinite dimensions the expectation 
value of the Hamiltonian~(\ref{1}) in terms of 
the Gutzwiller-correlated wave function~(\ref{GutzcorrdegbandsHund}) 
is given by 
\begin{mathletters}
\label{allresultsdegbandsHund}
\begin{eqnarray}
\langle \hat{H}\rangle 
&=&
\sum_{i\neq j;\bbox{\sigma_{1}},\bbox{\sigma'_{1}}}
\widetilde{t}_{i,j}^{\,\bbox{\sigma_{1}},\bbox{\sigma'_{1}}}
\langle 
\hat{c}_{i;\bbox{\sigma_{1}}}^{+}
\hat{c}_{j;\bbox{\sigma'_{1}}}^{\vphantom{+}}
\rangle_0
+\sum_{i;\bbox{\sigma}}\epsilon_{i;\bbox{\sigma}}n_{i;\bbox{\sigma}}
+\sum_{i;\Gamma }E_{i;\Gamma}m_{i;\Gamma }\;, \\[3pt]
\widetilde{t}_{i,j}^{\,\bbox{\sigma_{1}},\bbox{\sigma'_{1}}} 
&=&
\sum_{\bbox{\sigma_{2}},\bbox{\sigma'_{2}}}
{t}_{i,j}^{\bbox{\sigma_{2}},\bbox{\sigma'_{2}}}
\sqrt{q_{i;\bbox{\sigma_{2}}}^{\bbox{\sigma_{1}}}
q_{j;\bbox{\sigma'_{2}}}^{\bbox{\sigma'_{1}}}}\;.  \label{widetildet}
\end{eqnarray}\end{mathletters}% HERE!
It is seen that the variational ground-state energy can be cast into the
form of the expectation value of an effective single-particle Hamiltonian
with a renormalized electron transfer 
matrix~$\widetilde{t}_{i,j}^{\,\bbox{\sigma},\bbox{\sigma'}}$. 
Due to the off-diagonal terms in the local
interactions~(\ref{fullHat}) the $q$~factors are arranged in a non-diagonal
matrix $q_{i;\bbox{\sigma}}^{\bbox{\sigma'}}$ which determines the
quasi-particle bandwidth and the strength of band-mixing in the solid state.
As shown in the appendix, the elements of the
$\tensor{q}$~matrix can be written as 
\begin{eqnarray}
\sqrt{q_{\bbox{\sigma}}^{\bbox{\sigma'}}} 
&=& 
\sqrt{\frac{1}{n_{\bbox{\sigma'}}^{0}(1-n_{\bbox{\sigma'}}^{0})}} 
\sum_{\Gamma,\Gamma'} 
\sqrt{\frac{m_{\Gamma}m_{\Gamma'}}{m_{\Gamma}^0m_{\Gamma'}^0}}  
\nonumber \\[3pt]
&& \times 
\sum_{ {{\ I,I'} \atop {(\bbox{\sigma}\not\in I, \bbox{\sigma'}\not\in I')}}} 
\!\! \text{fsgn}(\bbox{\sigma'},I')\text{fsgn}(\bbox{\sigma},I) 
\sqrt{ m_{(I'\cup\bbox{\sigma'})}^{0} m_{I'}^{0} }
T_{\Gamma,(I\cup\bbox{\sigma})}^+ T_{(I'\cup\bbox{\sigma'}),\Gamma}
T_{\Gamma',I'}^+ T_{I,\Gamma'} \; ,  \label{qmatrix}
\end{eqnarray}
where we suppressed the site index and
used the definition~(\ref{deffsign}) of the fermionic sign function.

Eqs.~(\ref{allresultsdegbandsHund}) and~(\ref{qmatrix}) show that we may
replace the original variational parameters~$\lambda_{i;\Gamma }$ by their
physical counterparts, the atomic occupancies~$m_{i;\Gamma }$.
They are related by the simple equation 
\begin{equation}
\lambda_{i;\Gamma }^2=\frac{m_{i;\Gamma }}{m_{i;\Gamma }^0}\;.
\label{HundgutzrelationsA}
\end{equation}
Due to the local completeness relation the probability for an empty site is
a function of the other atomic occupation densities, 
\begin{equation}
m_{i;\emptyset }=1-\sum_{\Gamma \,(|\Gamma |=1)}m_{i;\Gamma }-
\sum_{\Gamma\,(|\Gamma |\geq 2)}m_{i;\Gamma }\;.  \label{completenessgamma}
\end{equation}
For the moment we suppress the site index. As shown in the appendix, the
parameters $\lambda_{\Gamma}^2$ for atomic configurations with a
single electron ($|\Gamma |=1$) are the eigenvalues of a $(2N)\times (2N)$
matrix~$\tensor{Z}$ whose entries are given by 
\begin{equation}
Z_{\bbox{\sigma},\bbox{\sigma'}}
=
\frac{n_{\bbox{\sigma}}^0}{m_{\bbox{\sigma}}^0}
\delta_{\bbox{\sigma},\bbox{\sigma'}}
-\sum_{\Gamma\, (|\Gamma |\geq 2)}\frac{m_{\Gamma} }{m_{\Gamma}^0}
\sum_{I\,(\bbox{\sigma},\bbox{\sigma'}\not\in I)}
\text{fsgn}(\bbox{\sigma'},I)\text{fsgn}(\bbox{\sigma},I)
T_{(I\cup \bbox{\sigma'}),\Gamma }T_{\Gamma ,(I\cup \bbox{\sigma})}^{+}
\frac{m_{I\cup (\bbox{\sigma},\bbox{\sigma'})}^0}%
{m_{(\bbox{\sigma},\bbox{\sigma'})}^0}\;.
\label{defZmatrix}
\end{equation}
The unphysical case $m_{\bbox{\sigma}}^0=0$ can safely be ignored. The
matrix $\tensor{Z}$ is diagonalized by a unitary matrix 
$\tensor{T}'$, 
\begin{equation}
(\tensor{T}')^{+}\tensor{Z}(\tensor{T}')=
\text{diag}(\lambda_{\Gamma}^2) \qquad  (|\Gamma|=1)
\end{equation}
with $T_{\bbox{\sigma},\Gamma}'=T_{\bbox{\sigma},\Gamma}^{*}$.
These entries in the matrix~$\tensor{T}$ remained undetermined at the end of
Sect.~\ref{Hamiltonians}.

Finally, the local densities $n_{i;\bbox{\sigma}}$ 
can be calculated from~(\ref{4c}) as 
\begin{mathletters}
\label{hh1}
\begin{equation}
n_{i;\bbox{\sigma}} = \sum_{I \, (\bbox{\sigma}\in I) } m_{i;I} \;,
\end{equation}
where the configuration probabilities for $|I|\geq 1$ 
follow from~(\ref{HundgutzrelationsCappendix}) of the appendix, 
\begin{equation}
m_{i;I} = \sum_{K} \biggl|
\sum_{\Gamma} 
\sqrt{\frac{m_{i;\Gamma}}{m_{i;\Gamma}^0}}
T_{i;\Gamma,I}^+ T_{i;K,\Gamma}
\biggr|^2 m_{i;K}^0 \;.  \label{mImGammastrich}
\end{equation}\end{mathletters}% HERE!
Hence, all quantities in~(\ref{allresultsdegbandsHund}) are now expressed in
terms of the variational parameters~$m_{i;\Gamma}$ and the properties 
of~$|\Phi_0\rangle$.

The remaining task is the minimization of $\langle \hat{H}\rangle$ 
in~(\ref{allresultsdegbandsHund}) with respect to 
$m_{i;\Gamma}$ and $|\Phi_0\rangle$. 
A conceivable though numerically unsuitable way to achieve this goal is
the following. For fixed $m_{i;\Gamma}$ an input wave 
function $|\Phi_0^{\text{in}}\rangle$ defines local 
occupancies $n_{i;\bbox{\sigma}}^{0,\text{in}}$. The wave function 
$|\Phi_0^{\text{out}}\rangle$ is the ground state of
the effective one-particle Hamiltonian 
\begin{mathletters}
\begin{eqnarray}
\hat{H}^{\text{eff, in}} &=& \sum_{i\neq j;\bbox{\sigma},\bbox{\sigma'}} 
\widetilde{t}_{i,j}^{\,\bbox{\sigma},\bbox{\sigma'};\text{in}} 
\hat{c}_{i;\bbox{\sigma}}^+
\hat{c}_{j;\bbox{\sigma'}}^{\vphantom{+}} 
+ \sum_{i;\bbox{\sigma}} 
\widetilde{\epsilon}_{i;\bbox{\sigma}}^{\, \text{in}} 
\hat{n}_{i;\bbox{\sigma}} 
+ \sum_{i;\Gamma} E_{i;\Gamma} m_{i;\Gamma} \label{Heffdef} \; , \\[3pt]
\widetilde{\epsilon}_{i;\bbox{\sigma}}^{\, \text{in}} &=&
\epsilon_{i;\bbox{\sigma}} 
\frac{n_{i;\bbox{\sigma}}^{\text{in}}}{n_{i;\bbox{\sigma}}^{0,\text{in}}}
\; . \label{epsilontildedef}
\end{eqnarray}\end{mathletters}% HERE
Note that we have to impose the condition that the orbitals do not mix locally
in our effective one-particle Hamiltonian,
$\widetilde{t}_{i,i}^{\bbox{\sigma},\bbox{\sigma'}}= 
\delta_{\bbox{\sigma},\bbox{\sigma'}}\widetilde{\epsilon}_{i;\bbox{\sigma}}$. 
The local occupancies of $|\Phi_0^{\text{out}}\rangle$ serve as input
for the next step in the iteration procedure. In this way the 
optimum $|\Phi_0^{\text{opt}}\rangle$ for fixed~$m_{i;\Gamma}$ 
is found recursively.
After the global minimization of~$\langle \hat{H}\rangle$ with respect to
the parameters $m_{i;\Gamma}$ the optimum effective one-particle Hamiltonian 
$\hat{H}^{\text{eff, opt}}$ defines a quasi-particle band structure which is
suitable for a comparison with experiments. Furthermore, it can be used
to derive the low-temperature thermodynamics.
Naturally, the application of $\hat{H}^{\text{eff, opt}}$ is restricted to 
the description of the low-energy physics 
(Fermi-liquid regime).~\cite{BUCH,GebhardPRB1,GebhardPRB2}

Another route to finite temperatures is the following.
The variational ground-state energy is a function of the
occupancies in momentum space.
Hence it can be used to derive Fermi liquid parameters~\cite{Vollhardtrev}
which give access to the low-energy physics of metallic
correlated-electron systems.~\cite{Nozieres,Leggett}

\subsection{Gutzwiller wave functions with pure density correlations}

If we ignore the non-diagonal
terms~${\cal J}_i^{\bbox{\sigma_1},\bbox{\sigma_2}; 
\bbox{\sigma_3},\bbox{\sigma_4}}$ in the atomic Hamiltonian~(\ref{fullHat})
we may set 
\begin{equation}
T_{I,\Gamma} \equiv \delta_{I,\Gamma}  \label{udiagonalapprox}
\end{equation}
in all the formulae of the previous subsection. Under these conditions we
recover the Gutzwiller wave functions with pure density correlations. 
In fact, for this class of Gutzwiller wave functions
the variational ground-state energy is independent of the non-diagonal 
terms~${\cal J}_i^{\bbox{\sigma_1},\bbox{\sigma_2}; 
\bbox{\sigma_3},\bbox{\sigma_4}}$.
This is ultimately due to
the fact that the correlator does not change the orbital occupation and that
Fock terms vanish in~$|\Phi_0\rangle$ according to~(\ref{noFOCKcops}).

Under the condition~(\ref{udiagonalapprox}) the $\tensor{Z}$~matrix 
in~(\ref{defZmatrix}) is diagonal. We use the fact that 
\begin{equation}
n_{i;\bbox{\sigma}} = m_{i;\bbox{\sigma}} 
+ \sum_{I\, (|I|\geq 2, \bbox{\sigma}\in I)} m_{i;I}  \label{msigmansigma}
\end{equation}
due to~(\ref{g2b}) so that the eigenvalues of $\tensor{Z}$ can be written as 
\begin{equation}
\lambda_{i;\bbox{\sigma}}^2 = 
\frac{n_{i;\bbox{\sigma}}^0 - n_{i;\bbox{\sigma}}+m_{i;\bbox{\sigma}}}%
{m_{i;\bbox{\sigma}}^0} \; .
\end{equation}
Hence, for consistency with~(\ref{HundgutzrelationsA}) we should have 
\begin{equation}
n_{i;\bbox{\sigma}} = n_{i;\bbox{\sigma}}^0  \label{densitiesagree}
\end{equation}
for Gutzwiller-correlated wave functions with pure density correlations.
This result can be derived more directly. As shown in the appendix, in
infinite dimensions we have 
\begin{equation}
\langle \hat{n}_{i;\bbox{\sigma}} \rangle = \langle \hat{P}_{i;\text{G}} 
\hat{n}_{i;\bbox{\sigma}} \hat{P}_{i;\text{G}} \rangle_0 = 
\langle \hat{n}_{i;\bbox{\sigma}} \hat{P}_{i;\text{G}}^2 \rangle_0 \; .
\end{equation}
For the second step we used the fact that now the Gutzwiller correlator
contains density operators only; compare~(\ref{GutzcorrdegbandsHund}). 
With the help of~(\ref{RHSb}) the result~(\ref{densitiesagree}) 
follows immediately. Thus, eqs.~(\ref{msigmansigma}) 
and~(\ref{densitiesagree}) allow us to express 
explicitly the probabilities for a
single occupancy in terms of the local densities in~$|\Phi_0\rangle$ and the
variational parameters $m_{i;I}$ for $|I|\geq 2$.

For pure density correlations and for wave functions which 
obey~(\ref{noFOCKcops}) different local configurations 
are not mixed. Consequently,
the $\tensor{q}$~matrix becomes diagonal, as can be shown 
explicitly from~(\ref{qmatrix}) with the help of~(\ref{udiagonalapprox}). 
For the diagonal elements we find 
\begin{mathletters}
\label{allresultsdegbands}
\begin{equation}
\sqrt{q_{i;\bbox{\sigma}}} \equiv \sqrt{q_{i;\bbox{\sigma}}^{\bbox{\sigma}}}
= \sqrt{\frac{1}{n_{i;\bbox{\sigma}}^0(1-n_{i;\bbox{\sigma}}^0)}} 
\sum_{I\, (\bbox{\sigma}\not\in I)} 
\sqrt{m_{i;I}m_{i;(I\cup\bbox{\sigma})}} \; ,
\end{equation}
and the expectation value of the Hamiltonian reduces to 
\begin{equation}
\langle \hat{H}\rangle = 
\sum_{i\neq j;\bbox{\sigma},\bbox{\sigma'}}
t_{i,j}^{\bbox{\sigma},\bbox{\sigma'}} 
\sqrt{q_{i;\bbox{\sigma}}} \sqrt{q_{j;\bbox{\sigma'}}} 
\langle 
\hat{c}_{i;\bbox{\sigma}}^+\hat{c}_{j;\bbox{\sigma'}}^{\vphantom{+}} 
\rangle_0 + \sum_{i;\bbox{\sigma}} \epsilon_{i;\bbox{\sigma}} 
n_{i;\bbox{\sigma}}^0 
+ \sum_{i;I} U_{i;I} m_{i;I} \; ,
\end{equation}
where 
\begin{equation}
U_{i;I} = \langle I |\hat{H}_{i;\text{at}} | I \rangle =
\sum_{\bbox{\sigma},\bbox{\sigma'}\in I} 
{\cal U}_{i}^{\bbox{\sigma},\bbox{\sigma'}} \;.
\end{equation}\end{mathletters}% HERE!

For translationally invariant systems the above 
equations~(\ref{allresultsdegbands}) were first derived 
in Refs.~\onlinecite{BWa,BWb} using
a generalized Gutzwiller approximation scheme. A concise description of this
semi-classical counting approach can be found 
in Ref.~\onlinecite{Vollhardtrev}; for a mathematically 
well-defined procedure, see Ref.~\onlinecite{JoergsZPB}. 
The wave function used in Refs.~\onlinecite{BWa,BWb,BGW} was defined as 
\begin{eqnarray}
|\Psi_{\text{G}}\rangle &=& \hat{P}_{\text{G}}' |\Psi_0\rangle 
\nonumber \\[3pt]
\hat{P}_{\text{G}}' &=& \prod_i
\prod_{I\, (|I|\geq 2)} 
g_{i;I}^{\hat{m}_{i;I}} \; .  \label{Gutzcorrdegbands}
\end{eqnarray}
The wave functions~$|\Psi_0\rangle$ and~$|\Phi_0\rangle$ are related by the
transformation~\cite{info2} 
\begin{mathletters}
\label{defpsizerophizero}
\begin{eqnarray}
|\Psi_0\rangle &=& \hat{P}_{\text{SP}} |\Phi_0\rangle \; , \\[3pt]
\hat{P}_{\text{SP}} &=& \prod_{i} g_{i;\emptyset} 
\prod_{\bbox{\sigma}=1}^{2N} g_{i;\bbox{\sigma}}^{\hat{n}_{i;\bbox{\sigma}}} 
\; .
\end{eqnarray}\end{mathletters}% HERE!
Since~$\hat{P}_{\text{SP}}$ contains single-particle operators only, 
both~$|\Psi_0\rangle$ and~$|\Phi_0\rangle$ are single-particle wave functions. 
The relation between the parameters~$g_{i;I}$ and $\lambda_{i;I}$ is given by 
\begin{mathletters}
\label{hh2}
\begin{eqnarray}
g_{i;\emptyset} &=& \lambda_{i;\emptyset} \; , \\[3pt]
g_{i;\bbox{\sigma}} &=& 
\frac{\lambda_{i;\bbox{\sigma}}}{\lambda_{i;\emptyset}} \; , \\[3pt]
g_{i;I} 
&=& 
\frac{\lambda_{i;I}\lambda_{i;\emptyset}^{|I|-1}}%
{\prod_{\bbox{\sigma}\in I}\lambda_{i;\bbox{\sigma}}} 
\quad (|I|\geq 2)\;,
\end{eqnarray}
and 
\begin{equation}
g_{i;I}^2 
= \frac{m_{i;\emptyset}^{|I|-1} m_{i;I} }{ \prod_{\bbox{\sigma}\in I} 
m_{i;\bbox{\sigma}}} \quad (|I|\geq 2)
\end{equation}\end{mathletters}% HERE!
holds in infinite dimensions.

Independently, the expressions~(\ref{allresultsdegbands}) were derived by
Hasegawa~\cite{Hasegawa} and Fr\'esard and Kotliar~\cite{Fresard} who used a
generalization of the Kotliar--Ruckenstein Slave-Boson mean-field 
approach~\cite{KRPRL} introduced by Dorin and Schlottmann.~\cite{Dorin} 
In Ref.~\onlinecite{BGW} we proved that the results of 
these approximate treatments
are variationally controlled in the limit of infinite dimensions; 
see Ref.~\onlinecite{BGW} for further comparison with previous variational and
Slave-Boson mean-field approaches to degenerate-band systems.

\section{Two degenerate bands}
\label{twodegbands}

The formulae derived in the appendix are completely general and apply for
all atomic Hamiltonians~(\ref{fullHat}) and for all kinds of symmetry
breaking in the one-particle wave function~$|\Phi_0\rangle $. Depending on
the complexity of the problem, the numerical treatment of multi-band
correlations may become rather involved. 
It appears to be a good strategy
to study the two-band case first which provides
the simplest example of a correlated multi-band model.
To keep our expressions for the $\tensor{q}$ and the 
$\tensor{Z}$~matrix as simple as possible, we chose a simple cubic
lattice with one atomic site per cell and two degenerate $d(e_g)$ orbitals
per atom. This model should reflect the situation of nickel to some
extent. For example, Ni$^{++}$, e.g. in NiO,
exhibits two $d(e_g)$ holes in a high spin state, and
metallic nickel has approximately one $d$~hole per site.

Alternatively, we could have chosen atoms with two $p_{(x,y)}$
orbitals on a square lattice. However, such a model might be less meaningful
for the study of ferromagnetic transitions since, as we will see below, 
these strongly depend on the structure of the
density of states which sensitively
depends on the dimension. In addition, our formulae become exact
for Gutzwiller-correlated wave functions in the limit of 
infinite dimensions, and $1/d$~corrections are expected to
be much smaller in three than in two dimensions. 
% NEW November 26, 1997
{}From our experience in the one-band case~\cite{METZNER,GebhardPRB1,Shiba}
we conjecture that the differences between $d=3$ and
$d=\infty$ are actually marginal.
% NEW November 26, 1997

\subsection{Atomic Hamiltonian}

We label the orbitals $d(3z^2-r^2)$ as $b=1$ and $d(x^2-y^2)$ as $b=2$, and
introduce the spin index $\sigma=\uparrow,\downarrow$. There are four
spin-orbitals per atom, leading to $2^4=16$ multi-electron configurations
(see table~\ref{tableone}). 
Then the atomic Hamiltonian reads
\begin{eqnarray}
\hat{H}_{\text{at}} &=&
U \sum_{b}\hat{n}_{b,\uparrow}\hat{n}_{b,\downarrow}
+U'\sum_{\sigma,\sigma'}\hat{n}_{1,\sigma}\hat{n}_{2,\sigma'}
-J\sum_{\sigma}\hat{n}_{1,\sigma}\hat{n}_{2,\sigma}
\label{twoorbhamiltonian} \\[3pt]
&& +J\sum_{\sigma}\hat{c}_{1,\sigma}^{+}
\hat{c}_{2,-\sigma}^{+}
\hat{c}_{1,-\sigma}^{\vphantom{+}}
\hat{c}_{2,\sigma}^{\vphantom{+}}
+J_{\text{C}} \Bigl(
\hat{c}_{1,\uparrow}^{+}\hat{c}_{1,\downarrow}^{+}
\hat{c}_{2,\downarrow}^{\vphantom{+}}\hat{c}_{2,\uparrow}^{\vphantom{+}}
+ 
\hat{c}_{2,\uparrow}^{+}\hat{c}_{2,\downarrow}^{+}
\hat{c}_{1,\downarrow}^{\vphantom{+}}\hat{c}_{1,\uparrow}^{\vphantom{+}}
\Bigr)\;.  \nonumber
\end{eqnarray}
For two orbitals, $\hat{H}_{\text{at}}$ exhausts all possible
two-body interaction terms.

All sixteen eigenstates and their respective energies
are given in table~\ref{tableone}.
The one-electron states and, due to the particle-hole symmetry, 
all three-electron states are seen to be degenerate.
The only non-trivial case are the two-electron states.
The model of two degenerate $d(e_g)$ orbitals leads to
the following restrictions enforced by symmetry:
first, as we can use real wave functions for $d(e_g)$ orbitals, 
the relation $J=J_{\text{C}}$ holds;
second, the relation $U-U'=2J$ follows from
the cubic symmetry.~\cite{Sugano}
To see this we address the six two-electron states.
There is one spin triplet ${}^3A_2$ 
with the energy $U'-J$. In addition, there are three
spin singlets: one, with symmetry ${}^1A_1$, 
has the energy $U+J_{\text{C}}$ whereas the other two have 
the energies $U'+J$ and $U-J_{\text{C}}$, respectively. 
Cubic symmetry requires~\cite{Sugano} that these two 
form the degenerate doublet $^1E$.
This symmetry requirement can be derived by a transformation 
into the equivalent basis $\left|3y^2-r^2\right\rangle $ 
and $\left| z^2-x^2\right\rangle $.

For Gutzwiller wave functions with pure density correlations
the exchange-type interactions (second line in~(\ref{twoorbhamiltonian}))
do not contribute to the variational 
ground-state energy. If we ignore these terms in~(\ref{twoorbhamiltonian})
all configurations~$|I\rangle$ are eigenstates of the resulting
$\hat{H}_{\text{at}}^{\text{dens}}$.
In addition to the states with zero, one, three, and four electrons
as listed in table~\ref{tableone}, the energies 
of the two-electron states are grouped into three doublets.
As listed in table~\ref{tabletwo}, there are 
(i)~the two components of the spin triplet with $\left| S_z\right| =1$ 
at the energy $U'-J$; 
(ii)~the states $\left| \uparrow ,\downarrow \right\rangle $ 
and $\left|\downarrow ,\uparrow \right\rangle $ at the energy 
$U'$, and (iii)~the states 
$\left| 0,\uparrow \downarrow \right\rangle $ and $\left| \uparrow
\downarrow ,0\right\rangle $ at the energy~$U$. 

\subsection{One-particle Hamiltonian and density of states}

We will use an orthogonal tight-binding Hamiltonian with first and second
nearest neighbor hopping matrix elements. Furthermore, we apply the two-center
approximation for the hopping matrix elements and exclude any spin-flip
hopping. Then the matrix elements in momentum space between the $3z^2-r^2$ 
($b=1$) and the $x^2-y^2$ ($b=2$) orbitals are given by~\cite{SlaterKoster} 
\begin{mathletters}
\label{baredispersions}
\begin{eqnarray}
\epsilon_1(k) 
&=&
t_{dd\sigma }^{(1)}((1/2)\cos k_x+(1/2)\cos k_y+2\cos k_z)
+(3/2)t_{dd\delta }^{(1)}(\cos k_x+\cos k_y)  \nonumber \\
&& +t_{dd\sigma }^{(2)}\cos k_x\cos k_y+[(1/4)t_{dd\sigma }^{(2)}
+3t_{dd\pi}^{(2)}](\cos k_x+\cos k_y)\cos k_z  \nonumber \\
&&+3t_{dd\delta }^{(2)}(\cos k_x\cos k_y+(1/4)\cos k_x\cos k_z
+(1/4)\cos k_y\cos k_z)\;, \\[3pt]
\epsilon_2(k) 
&=&
(3/2)t_{dd\sigma }^{(1)}(\cos k_x+\cos k_y)
+t_{dd\delta}^{(1)}((1/2)\cos k_x+(1/2)\cos k_y+2\cos k_z)  \nonumber \\
&&+4t_{dd\pi }^{(2)}\cos k_x\cos k_y+[(3/4)t_{dd\sigma }^{(2)}+t_{dd\pi}^{(2)}
+(9/4)t_{dd\delta }^{(2)}](\cos k_x+\cos k_y)\cos k_z\;,
\end{eqnarray}
and the band-mixing is given by 
\begin{eqnarray}
\epsilon_{12}(k) 
&=&
\epsilon_{21}(k)=(\sqrt{3}/2)[-t_{dd\sigma}^{(1)}
+t_{dd\delta }^{(1)}](\cos k_x-\cos k_y)  \nonumber \\
&&+[(\sqrt{3}/4)t_{dd\sigma }^{(2)}-\sqrt{3}t_{dd\pi }^{(2)}
+(3\sqrt{3}/4)t_{dd\delta }^{(2)}](\cos k_x-\cos k_y)\cos k_z\;.
\end{eqnarray}\end{mathletters}% HERE!
Here we put the cubic lattice constant equal to unity. As in 
Refs.~\onlinecite{BWa,BWb} the hopping parameters have been chosen according to
general experience for transition metal energy bands,
$t_{dd\sigma}^{(1)}=1\,$eV, $t_{dd\sigma }^{(2)}=0.25\,\text{eV}$, 
and $t_{dd\sigma}^{(1),(2)}:t_{dd\pi }^{(1),(2)}:t_{dd\delta }^{(1),(2)}=
1:(-0.3):0.1$. This
choice avoids pathological features in the energy bands such as perfect
nesting at half filling.

The one-particle part of the Hamiltonian~(\ref{1}) is easily diagonalized in
momentum space via the transformation 
\begin{mathletters}
\label{hh3}
\begin{eqnarray}
\eta_{k;1,\sigma }^{+} &=&\cos \phi_k\,\hat{c}_{k;1,\sigma }^{+}
+\sin \phi_k\,\hat{c}_{k;2,\sigma }^{+}\;, \\[3pt]
\eta_{k;2,\sigma }^{+} &=&-\sin \phi_k\,\hat{c}_{k;1,\sigma }^{+}
+\cos\phi_k\,\hat{c}_{k;2,\sigma }^{+}\;,
\end{eqnarray}
with 
\begin{equation}
\tan (2\phi_k)=\frac{2\epsilon_{12}(k)}{\epsilon_1(k)-\epsilon_2(k)}\;.
\end{equation}\end{mathletters}% HERE!
The dispersion relations for the $\eta $~bands become
\begin{equation}
E_{1,2}(k)=\frac{\epsilon_1(k)+\epsilon_2(k)}{2}
\pm \sqrt{\left( \frac{\epsilon_1(k)-\epsilon_2(k)}{2}\right) ^2
+[\epsilon_{12}(k)]^2\,}\;.
\label{Dispersiondiag}
\end{equation}
The $\eta $ bands are degenerate along the line $(\xi ,\xi ,\xi )$ in the
irreducible Brillouin zone, the total bandwidth is $W=6.6\,$eV. Our
one-particle state $|\Phi_0\rangle $ is chosen as 
\begin{equation}
|\Phi_0\rangle =\prod_{\sigma}
\prod_{{{k} \atop {(E_{1,2}(k)\leq E_{\text{F},\sigma })}}}
\hat{\eta}_{k;1,\sigma }^{+}\hat{\eta}_{k;2,\sigma }^{+}|\text{vacuum}\rangle 
\;.
\end{equation}
The Fermi surfaces of both $\eta $~bands are invariant under the symmetry
operations of the lattice.

The condition~(\ref{noFOCKcops}) is fulfilled due to the degeneracy of the 
$e_g$ orbitals. For the same reason the projected orbital densities of states 
\begin{eqnarray}
{\cal D}_1(\epsilon ) &=&\frac{1}{L}\sum_k
\cos ^2(\phi_k)\delta (\epsilon-E_1(k))
+\sin ^2(\phi_k)\delta (\epsilon -E_2(k))\;,  \nonumber \\[3pt]
{\cal D}_2(\epsilon ) &=&\frac{1}{L}\sum_k
\sin ^2(\phi_k)\delta (\epsilon-E_1(k))
+\cos ^2(\phi_k)\delta (\epsilon -E_2(k))\;,
\end{eqnarray}
have to be identical, ${\cal D}_1(\epsilon )={\cal D}_2(\epsilon )\equiv 
{\cal D}_0(\epsilon )/2$, and 
\begin{equation}
n_{b,\sigma }^0=\frac{1}{2}\int_{-\infty }^{E_{\text{F},\sigma }}d\epsilon 
{\cal D}_0(\epsilon )
\end{equation}
is independent of the band index, $n_{1,\sigma }^0=n_{2,\sigma }^0\equiv
n_{\sigma}^0$.

Since we built in the cubic symmetry into our starting wave function 
$|\Phi_0\rangle $ and our atomic Hamiltonian~(\ref{twoorbhamiltonian}) preserves
this symmetry, our selfconsistency cycle will not change this property.
Therefore, we may set $s_{1,\sigma }=s_{2,\sigma }\equiv s_{\sigma} $ and 
$t_{1,\sigma }=t_{2,\sigma }\equiv t_{\sigma} $ for our variational parameters;
compare table~\ref{tableone} for the notation. Note that the number of 
$\uparrow $~electrons and $\downarrow $~electrons need not be the same, i.e.,
we still allow for band ferromagnetism.

For the study of the ferromagnetic transition it is helpful to consider the
density of states at the Fermi energy, ${\cal D}_0(E_{\text{F},\sigma })$.
This quantity as a function of the band filling fraction $n_{\sigma} $ is
displayed in Fig.~\ref{DOSfig}. Later, we will study the half-filled case, 
$n_{\sigma} =0.5$, in the context of the Brinkman--Rice metal-to-insulator
transition, and the fillings $n_{\sigma} =0.29$
and $n_{\sigma} =0.35$ for ferromagnetism. The case $n_{\sigma} =0.29$
corresponds to a maximum in the density of states at the Fermi energy.
There we expect the strongest tendency to ferromagnetism.

In this work we take the viewpoint of the canonical rather than the
grand-canonical ensemble. This means that we keep the zero of energy fixed
for all band fillings. Then, the Fermi energy moves as a function of the
band filling and is different for the two spin species in the case of
ferromagnetism. Of course, this does not change the results because we could
have kept the Fermi energy the same for both bands and shifted the minority
band against the majority band to vary the magnetization density.

\subsection{Variational ground-state energy}
\label{groundstateenergysect}

Now we derive the explicit form of the ground-state energy 
functional~(\ref{allresultsdegbandsHund}) for our example. 
To this end we first show that
the matrix $\tensor{Z}$ in~(\ref{defZmatrix}) is diagonal. From 
table~\ref{tableone} we see that for $|\Gamma|> 2$ the
atomic eigenstates are also configuration eigenstates, 
$T_{I,\Gamma}=\delta_{I,\Gamma}$. In this case, 
the factors $T_{(J\cup\bbox{\sigma}),\Gamma}
T_{\Gamma,(J\cup\bbox{\sigma'})}^+$ require $J\cup\bbox{\sigma}=I
=J\cup\bbox{\sigma'}$, i.e., $\bbox{\sigma}= \bbox{\sigma'}$. For 
$|\Gamma|=2$ we note that $J=\bbox{\gamma}$ with $\bbox{\gamma}\neq 
\bbox{\sigma},\bbox{\sigma'}$ has to hold. Now that $|\Gamma\rangle=
\sqrt{1/2}[|\bbox{\sigma_1},\bbox{\sigma_2}\rangle \pm 
|\bbox{\sigma_3},\bbox{\sigma_4}\rangle]$ 
according to table~\ref{tableone} we see again that 
$\bbox{\sigma}=\bbox{\sigma'}$ must hold since either 
$(\bbox{\gamma},\bbox{\sigma}) = (\bbox{\sigma_1},\bbox{\sigma_2}) 
= (\bbox{\gamma},\bbox{\sigma'})$ or $(\bbox{\gamma},\bbox{\sigma}) 
= (\bbox{\sigma_3},\bbox{\sigma_4}) = (\bbox{\gamma},\bbox{\sigma'})$ 
must be fulfilled ($(\bbox{\sigma_1},\bbox{\sigma_2}) \cap 
(\bbox{\sigma_3},\bbox{\sigma_4})=\emptyset$). Since the $\tensor{Z}$~matrix 
is diagonal it follows that the
matrix $T_{\bbox{\sigma},\Gamma}'$ is the unit matrix.

The eigenvalues of the $\tensor{Z}$~matrix are 
$\lambda_{\bbox{\sigma}}^2=m_{\bbox{\sigma}}/m_{\bbox{\sigma}}^0$. 
Then, the relation
\begin{equation}
m_{\Gamma} ^0=m_I^0  \label{mgammanulminull}
\end{equation}
is fulfilled for all $\Gamma $, $I$ with $|\Gamma |=|I|$. 
We then find from~(\ref{completenessgamma}) that 
\begin{mathletters}
\label{hh4}
\begin{equation}
s_{\sigma} =n_{\sigma}^0-\left[ d_t^{\sigma \sigma }+t_{-\sigma }+2t_{\sigma}
+f +\frac{1}{2}\left( d_A+2d_E+d_t^0\right) \right]
\end{equation}
gives the probabilities for a single occupancy in terms of the multiple
occupancies which serve as our variational parameters; see table~\ref{tableone}
for the notation. 
The probability for an empty site follows from the completeness
relation~(\ref{localcompelete}) as 
\begin{equation}
e=1-2n_{\uparrow }^0-2n_{\downarrow }^0+d_t^{\uparrow \uparrow}+
d_t^{\downarrow \downarrow }+d_t^0+d_A+2d_E+4t_{\uparrow }+
4t_{\downarrow}+3f\;.
\end{equation}\end{mathletters}% HERE!
Along the same lines it can be shown that eq.~(\ref{mImGammastrich}) reduces
to 
\begin{mathletters}
\label{hh5}
\begin{equation}
m_I=\sum_{\Gamma} |T_{I,\Gamma }|^2m_{\Gamma} \;.  \label{mImGammasimple}
\end{equation}
With the help of~(\ref{defZmatrix}) and~(\ref{mgammanulminull}) it then
follows that 
\begin{equation}
n_{\sigma} =n_{\sigma}^0
\end{equation}
holds in our model. Similar arguments, as were used to show that 
the $\tensor{Z}$ matrix is diagonal, 
can be employed to show that the matrix $\tensor{q}$
is diagonal. Furthermore, the degeneracy of the orbitals due to the
cubic symmetry requires 
\begin{equation}
q_{\bbox{\sigma}}^{\bbox{\sigma}}\equiv q_{\sigma} \; .
\end{equation}\end{mathletters}% HERE!
Hence, the dispersion relations of the $\eta $~bands are rescaled by the
same factor~$q$ such that $|\Phi_0\rangle $ is unchanged, and our
self-consistency cycle terminates after a single iteration. Thus the optimum 
$|\Phi_0\rangle $ can be chosen from the start. Nevertheless, we still
allow for ferromagnetism since $E_{\text{F},\sigma }$ remained undetermined
thus far.

A straightforward calculation gives the explicit form of the $q$~factors, 
\begin{eqnarray}
q_{\sigma}
&=&
\frac{1}{n_{\sigma}^0(1-n_{\sigma}^0)}
\biggl[ \left( 
\sqrt{t_{\sigma} }+\sqrt{s_{-\sigma }}\right) 
\frac{1}{2}\left( \sqrt{d_A}+2\sqrt{d_E}+\sqrt{d_t^0}\right)  
\nonumber \\[3pt]
&&\hphantom{\frac{1}{n_{\sigma}^0(1-n_{\sigma}^0)} \Bigl[ }
+\sqrt{s_{\sigma} }
\left( \sqrt{e}+\sqrt{d_t^{\sigma \sigma }}\right) 
+\sqrt{t_{-\sigma }}
\biggl( \sqrt{d_t^{(-\sigma)(-\sigma) }}+\sqrt{f}\biggr) 
\biggr]^2\;.
\end{eqnarray}
We denote the kinetic energy of the $(1,\sigma )$ and $(2,\sigma )$
electrons in the uncorrelated state~$|\Phi_0\rangle $ by 
\begin{equation}
\overline{\epsilon }_{\sigma ,0}=\int_{-\infty }^{E_{\text{F},\sigma}}
d\epsilon \,\epsilon {\cal D}_0(\epsilon )\;.
\end{equation}
With the help of table~\ref{tableone} we may then cast the minimization
problem into the form 
\begin{eqnarray}
E^{\text{var, atom}} &=&
\sum_{\sigma} q_{\sigma} \overline{\epsilon }_{\sigma,0}
+(U'-J)(d_t^{\uparrow \uparrow }+d_t^{\downarrow \downarrow}+d_t^0)  
\nonumber \\[3pt]
&&+2(U'+J)d_E+(U+J)d_A+(2U+4U'-2J)(t_{\uparrow}+t_{\downarrow }+f)\;.
\end{eqnarray}
This expression must be minimized with respect to the eight variational
parameters, $d_t^{\uparrow \uparrow }$, $d_t^{\downarrow \downarrow }$, 
$d_t^0$, $d_t^{\text{0}}$, $d_E$, $d_A$, $t_{\uparrow }$, $t_{\downarrow }$,
and $f$ for a given band-filling~$n_{\sigma} $. In a paramagnetic situation,
$n_{\uparrow}=n_{\downarrow}$,
the number of variational parameters is reduced to five by the relations 
$d_t^{\sigma \sigma }=d_t^{\text{0}}\equiv d_{\text{t}}$, 
$t_{\sigma} \equiv t$, and $q_{\sigma} =q$. Furthermore, 
the relations $s_{\uparrow }=s_{\downarrow }$ and 
$\overline{\epsilon}_{\downarrow ,0}=\overline{\epsilon }_{\uparrow ,0}$
hold.

For Gutzwiller wave functions with density correlations we employ 
eqs.~(\ref{allresultsdegbands}) and the notations of table~\ref{tabletwo}. 
Now the variational problem reads 
\begin{mathletters}
\label{twobandsnohund}
\begin{eqnarray}
E^{\text{var, dens}} &=&\sum_{\sigma} \widetilde{q}_{\sigma} 
\overline{\epsilon }_{\sigma ,0}+(U'-J)(d_1^{\uparrow \uparrow}
+d_1^{\downarrow \downarrow })  \nonumber \\[3pt]
&&+2U'd_s+2Ud_c+(2U+4U'-2J)(t_{\uparrow }+t_{\downarrow}+f)\;.
\end{eqnarray}
Here the $q$~factors are given by 
\begin{eqnarray}
\widetilde{q}_{\sigma} &=&
\frac{1}{n_{\sigma}^0(1-n_{\sigma}^0)}
\biggl[ 
\left( \sqrt{t_{\sigma}}+\sqrt{s_{-\sigma }}\right) 
\left( \sqrt{d_c}+\sqrt{d_s}\right)  
\nonumber \\[3pt]
&&\hphantom{\frac{1}{n_{\sigma}^0(1-n_{\sigma}^0)} \Bigl[ }
+\sqrt{s_{\sigma} }\left( \sqrt{e}+\sqrt{d_1^{\sigma \sigma }}\right) 
+\sqrt{t_{-\sigma }}\biggr( \sqrt{d_1^{(-\sigma )(-\sigma )}}+\sqrt{f}\biggr) 
\biggr]^2\;.
\end{eqnarray}\end{mathletters}% HERE!
In this case our variational parameters are $d_1^{\uparrow \uparrow }$, 
$d_1^{\downarrow \downarrow }$, $d_s$, $d_c$, $t_{\uparrow }$, $t_{\downarrow}$,
and $f$. The probabilities for an empty site~$e$ and a singly occupied
site~$s_{\sigma} $ are related to the variational parameters by 
\begin{mathletters}
\label{hh6}
\begin{eqnarray}
s_{\sigma} &=&n_{\sigma}^0-\left[ d_1^{\sigma \sigma }+t_{-\sigma }+2t_{\sigma}
+f+d_c+d_s\right] \;,  \\[3pt]
e &=&1-2n_{\uparrow }^0-2n_{\downarrow }^0+d_1^{\uparrow \uparrow}
+d_1^{\downarrow \downarrow }+2d_s+2d_c+4t_{\uparrow }+4t_{\downarrow}+3f\;.
\end{eqnarray}\end{mathletters}% HERE!
The expression~(\ref{twobandsnohund}) is identical to the one used in 
Refs.~\onlinecite{BWa,BWb,Hasegawa}.

\subsection{Brinkman--Rice metal--insulator transition at half band-filling}

As a first application of our variational treatment we study the
Brinkman--Rice metal-to-insulator transition. 
For a single band and a translationally invariant system
we recover the original Gutzwiller wave function.~\cite{Gutzwiller1963}
In infinite dimensions this wave function at half band-filling is known to
describe a continuous transition from the paramagnetic metal 
to a paramagnetic insulator at $U=U_{\text{BR}}$ above
which all electrons are localized 
(Brinkman--Rice insulator).~\cite{BrinkmanRice}
It can be shown, though, that the Brinkman--Rice transition 
at a finite interaction strength is the consequence
of the large-$d$ limit, i.e., it is not contained
in the wave function for any finite dimension.~\cite{PvDFloriVollhardt}
Hence, statements on the metal-insulator transition based
on our variational description must be taken with care.
Even in infinite dimensions the Brinkman--Rice transition
can be concealed by an (antiferromagnetically) 
ordered phase; see Ref.~\onlinecite{Fazekas} for the one-band case and
Ref.~\onlinecite{Hasegawa} for $N=2$. 
It should be clear, though, that the onset of
long-range order crucially depends
on the choice of the matrix elements for the electron transfer.
In general, there is no perfect nesting between the Fermi surface
and the Brillouin zone such that antiferromagnetism
is not expected to set in for small interaction strengths.

For multi-band systems the Brinkman--Rice transition occurs 
at integer numbers~$1\leq n\leq 2N-1$ of electrons per atom. 
For two bands the transitions for $n=1$ or $n=3$ 
are continuous like in the one-band case. 
There, our results do not differ much from those given in 
Ref.~\onlinecite{BWa} for Gutzwiller wave functions with
pure density correlations.
Thus we focus on the case $n=2$, where, in general, the transition is
{\em discontinuous\/} in the bandwidth reduction factor~$q$. 
This means that a jump occurs at the Brinkman--Rice transition from 
the finite value $q_{\text{BR}}$ 
in the metallic phase to $q=0$ in the insulating phase. 

For $n=2$ the dependence of the five variational parameters $d_t$, $d_E$,
$d_A$, $t$, $f$ as a function of the interaction strength~$U$ 
is shown in Fig.~\ref{varparameter}. Here,
the value $J=0.2\,U$ ($U'=0.6\,U$) was chosen.
For $U=0$ (independent electrons) the values of all quantities are equal.
As $U$~becomes larger the spin-triplet double occupancy $d_t$ increasingly
dominates the other multiple occupancies; in particular, this is true
close to the jump at $U_{\text{BR}}$. All multiple occupancies 
are discontinuous at the Brinkman--Rice transition
since, in the insulating case, all electrons are frozen into local spin
triplets, i.e., we have $d_t=1/3$ and all other multiple occupancies are 
zero. Note that in the case of pure density
correlations the dominance of $d_t$ is less pronounced; compare Fig.~2 of 
Ref.~\onlinecite{BWa}.

In Fig.~\ref{qfuncUJ} the $q$ values are shown as a function of $U$ for
various $J/U$ ratios. The singular case $J=0$ ($U=U'$)
differs from the generic situation both qualitatively
and quantitatively. The Brinkman--Rice transition is continuous {\em only\/}
at this point, and values of $J/U$ as small
as $10^{-2}$ produce finite jumps of a significant size.
A (realistic) value of $U'=0.8U$ ($J=0.1U$) is enough to 
reduce the critical interaction strength $U_{\text{BR}}$ 
for the Brinkman--Rice transition by a factor of two; see Fig.~\ref{qfuncUJ}.
Only for $U'=U$ ($J=0$) all atomic two-electron states 
are degenerate in energy. Thus, near the Brinkman--Rice transition, 
all double occupancies have equal weight, both in the metallic and in 
the insulating phase. Any finite~$J$ value
will remove the degeneracies and reestablish the generic case. At the
singular point of `zero configuration width'~\cite{Hubbard}
the critical interaction
strength can be given by an analytical expression, first 
derived by Lu.~\cite{Lu}

As seen from Fig.~\ref{qfuncUJ} the critical interaction
strength~$U_{\text{BR}}$ and the size of the $q$~factor
strongly depend on the size of the Hund's-rule coupling~$J/U$.
In Fig.~\ref{qdiscontinuity} we display the behavior of $q_{\text{BR}}$ 
as a function of $U'/U$ 
at the corresponding critical interaction strengths $U_{\text{BR}}$.
For the Gutzwiller wave function with atomic correlations (GW$_{\text{atom}}$)
a maximum of $q_{\text{BR}}^{\text{max}}\approx 0.4$ for $U'<U$ 
occurs near $U'/U=0.9$ ($J/U=0.05$). 
A shallow minimum of $q_{\text{BR}}^{\text{min}}\approx 0.1$ is seen
near $U'/U\approx 0.14$ ($J/U=0.43$). 
On the contrary, the same curve for the Gutzwiller wave function
with pure density correlations (GW$_{\text{dens}}$) increases monotonically as
a function of~$J/U$ towards $q_{\text{BR}}^{\text{dens}}\approx 0.6$ 
at $U'/U=0$. 

In the range $0\leq U'/U<1$ 
we always find $q_{\text{BR}}^{\text{dens}}>q_{\text{BR}}$.
Moreover, we have \hbox{$U_{\text{BR}}>U_{\text{BR}}^{\text{dens}}$},
see Fig.~\ref{phasedia}.
As expected, the metallic state is stabilized by the introduction of
the full atomic correlations. Nevertheless, the two values
for the Brinkman--Rice transition are fairly close to each other,
$U_{\text{BR}}\gtrsim U_{\text{BR}}^{\text{dens}}$.
Therefore, it is interesting to plot the value of the $q$~factor
for the case of atomic correlations at $U=U_{\text{BR}}^{\text{dens}}$.
It is very similar to the $q_{\text{BR}}$ curve for pure density correlations
(see Fig.~\ref{qdiscontinuity}). This
shows that the $q$~factor sharply --yet continuously--
drops as a function of~$U$
in the region $U_{\text{BR}}^{\text{dens}}\leq U <
U_{\text{BR}}$ before it jumps from $q_{\text{BR}}$ to zero at
the Brinkman--Rice transition.

The Brinkman--Rice transition is discontinuous because the ``metallic''
and the ``insulating'' minimum compete for the global minimum
of the variational energy function. In contrast to the one-parameter
minimization problem of the single-band case (and the two-band case
for $J=0$) the metallic minimum does not smoothly develop into the
insulating one in the presence of more than one atomic energy scale.
The size of the $q$-factor jump
measures the difference between the variational
parameters in the metallic and in the insulating phase.
Small discontinuities imply that the variational parameters
of the metallic state at the transition are close to those
of the Brinkman--Rice insulator.
For large $q_{\text{BR}}$ the metallic and the insulating minimum
are well separated in parameter space.

In the Brinkman--Rice insulator all sites are in the state with lowest
atomic energy. In the metal higher atomic states are mixed in.
The strength of the mixing depends on the energy separation
between the atomic levels. For example, the three 
$S=1$ configurations at energy $U-3J$ are separated from the 
singlet $^1E$ at energy $U-J$ by~$2J$. 
Therefore, the $S=1$ configurations dominate the metallic state
near the Brinkman--Rice transition for large~$J$, and 
$q_{\text{BR}}$ decreases with increasing~$J$. 
When $J$ becomes too large, near $U'=0$ ($J/U=0.5$), the energy of
the three-electron states $3U-5J$ is approaching the energies of the
two-electron states. Thus, the value of $t$ is enhanced 
at the expense of the $d_t$ parameter. 
Consequently, $q_{\text{BR}}$ increases again.
In the case of pure density correlations, the two $\left| S_z\right| =0$
configurations $\left| \uparrow ,\uparrow \right\rangle $ and $\left|
\downarrow ,\downarrow \right\rangle $ have energies $U-3J$ not too much
lower than the two configurations $\left| \uparrow ,\downarrow \right\rangle 
$ and $\left| \downarrow ,\uparrow \right\rangle $ with energies $U-2J$.
This leads to a competition of the respective occupancies for all~$J$,
and $q_{\text{BR}}^{\text{dens}}$ is a fairly smooth function of~$J$.

In Fig.~\ref{phasedia} we display the paramagnetic $(U,U')$~phase
diagram at half band-filling. It is seen that the additional atomic
correlations (GW$_{\text{atom}}$) stabilize 
the metallic phase for all $U>U'$
($J>0$) compared to the result of the density correlations. 
The figure also shows the gain in
the variational energy when we use the Gutzwiller wave functions
with full atomic correlations instead of pure density correlations.
The gain is shown for fixed value $U=U_{\text{BR}}^{\text{dens}}$ 
as a function of $U'/U$.
It is quite considerable, of the order of $0.1\,$eV,
for realistic values of $J/U\approx 0.1$. 

For $J<0$ ($U'>U$), the metal is {\em less\/} stable
in the presence of full atomic correlations.
Note that in this parameter range the
insulating ground state is different for the two variational
wave functions: a unique atomic $^1A_1$ state with 
energy $U-\left| J\right| $
versus two degenerate $\left| \uparrow \downarrow,0\right\rangle $ 
and $\left| 0,\uparrow \downarrow \right\rangle $ states
of energy $U$ for pure density correlations. 
As a consequence, in this limit the violation of the atomic symmetry
leads to a qualitatively different result for
pure density correlations.

Finally, in Fig.~\ref{Ssquared}, we display the size of the local spin 
$\langle \bigl(\vec{S}_i\bigr)^2\rangle =S_i(S_i+1)$ at half band-filling. In
the Brinkman--Rice insulator we have $S_i(S_i+1)=2$ ($S_i=1$) when $J>0$.
For $J<0$ the local spin drops to zero at $U_{\text{BR}}$ 
since the singlet $^1A_1$ is the local ground-state configuration
in the Brinkman--Rice insulator. We again focus on $J\geq 0$.
For non-interacting electrons ($U=0$) simple statistical arguments apply
and the local spin is readily found to be 
$(1/2)(1+1/2)(8/16)+1(1+1)(3/16)=3/4$.

For $J=0$ the local spin increases very slowly with $U$, up to 
$1(1+1)(3/6)=1$ at $U_{\text{BR}}$ and above. Recall that for~$J=0$
the Brinkman--Rice insulator is highly degenerate. 
For $J>0$ the weight of the local triplets becomes
more and more important towards the Brinkman--Rice transition.
This leads to local spin values for $U=U_{\text{BR}}$ as
large as $S_i(S_i+1)=1.55$ for $J/U=0.1$ and even 
$S_i(S_i+1)=1.8$ for $J/U=0.45$. As seen above in Fig.~\ref{qdiscontinuity},
the increase in the local spin is most prominent 
close to the Brinkman--Rice transition which again demonstrates
the drastic change in the multiple occupancies there.

\subsection{Itinerant ferromagnetism}

The formulae which we derived in Sect.~\ref{groundstateenergysect} equally
apply for the case of ferromagnetism. In this subsection we allow for a
finite magnetization density $M$ per band in the $z$~direction, 
\begin{equation}
0\leq M=(n_{b,\uparrow }-n_{b,\downarrow })/2\leq M_{\text{sat}}=n/4\;.
\end{equation}
In Fig.~\ref{magnetizationdensity}, the magnetization~$M$ is shown 
as a function of $U$ for fixed $J/U=0.2$ ($U'/U=0.6$). 
The critical interaction for the ferromagnetic transition, 
$U_{\text{F}}^{\text{atom}}$, is
about a factor two larger than its value 
$U_{\text{F}}^{\text{HF}}$ obtained from
the Hartree--Fock--Stoner theory. The corresponding 
values $U_{\text{F}}^{\text{dens}}$ always
lie somewhat below the values for the
Gutzwiller wave function with atomic correlations.
In general, the relation $M_{\text{HF}}(U)> 
M_{\text{dens}}(U)> M_{\text{atom}}(U)$ holds, i.e.,
for all interaction strengths
the tendency to ferromagnetism is strongest within the Hartree--Fock
theory and weakest for Gutzwiller wave functions with atomic correlations.
Furthermore, the slopes of $M(U)$ are
much steeper in the Hartree--Fock results than in the presence of
correlations. 

The properties of the ferromagnetic phase
strongly depend on the spectrum of the atomic two-electron
configurations. To further analyze this point we included
the case of $J_{\text{C}}=0$, which changes only 
the excited two-electron states. A shift of the curve $M(U)$ results
towards smaller interaction strengths; for a given
magnetization density a smaller interaction strength is required as
compared to the correct symmetry case $J=J_{\text{C}}$
(see Fig.~\ref{magnetizationdensity}).
The effect is more pronounced when we go to the Gutzwiller wave function
with pure density correlations. In this case
all exchange terms in~(\ref{twoorbhamiltonian}) are neglected.
Then, even the ground state is modified since
the atomic spin triplet with $S^z=0$ moves up in energy into the range
of the atomic spin singlets. Again, the magnetization curve
shifts to (much) smaller interaction strengths.
Both results indicate how strongly itinerant ferromagnetism
is influenced by the atomic ${\frak n}$-electron spectra.

In Fig.~\ref{magnetizationdensity}a 
we chose the particle density per band to be
$n/4=0.29$ (more precisely: $n/4=0.2941$), right at
the maximum of the density of state curve; compare Fig.~\ref{DOSfig}.
For this case there are finite slopes of the $M(U)$ curves at $U_{\text{F}}$, 
and a ``Stoner criterion'' for the onset of ferromagnetism
applies. In Fig.~\ref{magnetizationdensity}b we chose
the particle density per band as $n/4=0.35$. 
As seen from the density of states
in Fig.~\ref{DOSfig}, the density of states at the Fermi energy 
${\cal D}_0(E_{\text{F},\uparrow})+{\cal D}_0(E_{\text{F},\downarrow })$ 
first {\em increases\/} as a function of the magnetization density, and,
therefore, a discontinuous transition occurs from the paramagnet to
the ferromagnet.

In the case of pure density correlations a second jump in the $M(U)$ curve is
observed, which is absent in the other two curves. 
As was discussed in Ref.~\onlinecite{BWb}, 
this jump is related to another feature of the density of states. In the
Hartree--Fock theory this feature is too weak to be of any significance 
in comparison to the interaction energy. When the full 
atomic correlations are taken into account, this 
first-order jump at a finite magnetization density
disappears due to the enhanced flexibility of the 
variational wave function. Nevertheless, in this range of 
a strongly varying magnetization density
we find rapid variations of the various double
occupancies, similar to the behavior near the 
Brinkman--Rice transition for~$n/4=0.5$.

Another remarkable difference between the Hartree--Fock and the Gutzwiller
method lies in the approach to ferromagnetic
saturation.
In the Hartree--Fock theory the magnetization saturates
at $U$~values about 20\% to 40\% above
the onset of ferromagnetism at $U_{\text{F}}^{\text{HF}}$. 
In contrast, in the variational approach saturation is reached
at about twice the onset value, $U_{\text{sat}}\lesssim 2U_{\text{F}}$. 
However, even when the minority spin occupancies are zero
and $\langle \hat{S}_z^{\text{at}} \rangle$ is constant, the majority spin
occupancies $s_{\uparrow }$ and $d_t^{\uparrow \uparrow }$ vary 
with~$U$ since the limit of zero empty sites is reached
only for $U\to\infty$.

The magnitude of the local spin as a function of~$U$ is shown
in Fig.~\ref{Localspinferro}.
For $U\to\infty$ each site is either singly 
occupied with probability $2-n$ or doubly occupied (spin $S=1$) 
with probability $n-1$. Hence, 
$\langle (\vec{S}_i)^2\rangle_{\infty} =(3/4)(2-n)+2(n-1)=
5(n/4)-1/2$. For the correlated wave functions 
this limit is reached from {\em above\/} since, for $U<\infty$, charge
fluctuations first increase the number of spin-one sites at the expense of
spin-$1/2$ sites which turn into empty sites. A further decrease of $U$ will
also activate the singlet double occupancies and higher multiple
occupancies. Thus, the local spin eventually reduces
below $\langle (\vec{S}_i)^2\rangle_{\infty}$. 
On the contrary, Hartree--Fock theory
does not give the proper large-$U$ limit for the local spin.
Instead, the Hartree--Fock limit is
given by $\langle (\vec{S}_i)^2\rangle_{\infty}^{\text{HF}}
=(n/4)(3+n/2)$.

The change of $\langle (\vec{S}_i)^2\rangle $ 
at $U_{\text{F}}$ is only a minor effect within the 
correlated-electron approach. In particular, this holds
for the case of atomic correlations, where about 90\% of the local spin 
saturation value is already reached in the paramagnetic state.
Again, the Hartree--Fock results are completely different.
There, the local spin sharply increases as a function of
the interaction strength since the absence of correlations fixes
$\langle (\vec{S}_i)^2\rangle^{\text{HF}}(U< U_{\text{F}}^{\text{HF}})
= \langle (\vec{S}_i)^2\rangle(U=0)$.

In Fig.~\ref{Phasediaferro} we display the $J$-$U$ phase diagram for both
fillings. It shows that Hartree--Fock theory always predicts a ferromagnetic
instability. In contrast, the correlated-electron approach strongly supports
the idea that a substantial on-site exchange is required for the occurrence
of ferromagnetism at realistic interaction strengths. For the case
$n/4=0.29$, the differences between the phase diagrams for the two
correlated-electron wave functions are minor. Due to the large density
of states at the Fermi energy, the critical interaction strengths 
for the ferromagnetic transition are comparably small, 
and the densities for the double occupancies in both
correlated wave functions do not differ much. 
For the larger band filling $n/4=0.35$, i.e., away from 
the peak in the density of state, 
the values for $U_{\text{F}}$ are larger and, in the
atomic correlation case, the Gutzwiller wave functions can more easily
generate local spin triplets while keeping the global paramagnetic phase.

Finally, in Fig.~\ref{condenenergy}, we display the energy differences
between the paramagnetic and ferromagnetic ground states as a function of
the interaction strength for $J=0.1U$. For the correlated-electron case
this quantity is of the order of the Curie temperature which is in the
range of $100\,\text{K}-1000\,\text{K}$ in real materials. 
On the other hand, the
Hartree--Fock theory yields small condensation energies only in the range 
of $U\approx 4\,$eV; for larger $U$, the condensation energy 
is of order $U$.
Including the correlation effects we have
relatively small condensation energies even for interaction values
as large as twice the bandwidth ($U\approx10\,$eV).

\section{Summary and Conclusions}
\label{conclusions}

In this work we constructed Gutzwiller-correlated wave functions with atomic
correlations for general multi-band Hubbard models. 
We evaluated these many-particle wave functions in the limit of infinite
space dimensions ($d=\infty$) for the general atomic Hamiltonian for all
interaction strengths without any restrictions on the electron transfer
matrix elements between orbitals on the same or on different lattice sites. 
% NEW November 26, 1997
Within the metallic phase the differences between three and infinite
dimensions were found to be small for the
Gutzwiller wave function for a single band.~\cite{METZNER,GebhardPRB1,Shiba} 
% NEW November 26, 1997
Therefore, we expect that our results are well applicable for the case of 
physical interest.

Our variational states consist of a Jastrow-type many-particle
correlator which acts on an appropriate Hartree--Fock single-particle
product wave function. The Gutzwiller correlator is chosen to modify 
the occurrence of atomic multi-electron eigenstates~$|\Gamma_i\rangle$
as compared to the uncorrelated (statistical) case.
Therefore, our trial states are exact both in the non-interacting 
and in the atomic limit, and they incorporate the essential competition
between local and itinerant features of interacting multi-band systems.

The atomic single-particle states of appropriate symmetry (spin-orbits)
constitute the basis for our one-electron Hamiltonian which 
describes the motion of the electrons through the solid and
provides the Hartree--Fock wave function. The atomic multi-electron
configurations $|I_i\rangle$ are product states (`Slater determinants')
made from the spin-orbit states. 
If the (on-site) electron--electron interaction
contains only density-type two-particle interactions, the configurations 
$|I_i\rangle$ will not couple, and the local probabilities $m_{i;I}$ of 
these ${\frak n}$-electron configurations (${\frak n}=|I|$) can be
used as variational parameters to minimize the ground-state energy.
In general, however, the states $|I_i\rangle$ do not exhibit the correct
symmetry of atomic ${\frak n}$-electron eigenstates~$|\Gamma_i\rangle$. 
The correct symmetry can be established only when all exchange-type 
terms of the atomic Hamiltonian are taken into account, i.e.,
all atomic correlations must be included from the beginning.
Hence, all configurations $|I_i\rangle$
within the subspace $|I_i|={\frak n}$ are coupled, and
the diagonalization of the atomic Hamiltonian
by unitary matrices $T_{i;I,\Gamma}$ results in the 
${\frak n}$-electron atomic eigenstates $|\Gamma_i\rangle$.
Therefore, the multiple occupancies $m_{i;\Gamma}$ for the atomic eigenstates
are the appropriate variational parameters in our problem. In many
cases the elements of the matrices $\tensor{T}$ are given by symmetry alone; 
in general, however, they must be obtained from the
diagonalization of the atomic Hamiltonian.

The exact results in infinite dimensions can be cast into the form
of an {\em effective\/} single-particle Hamiltonian with reduced
electron transfers between the lattice sites.
Since the atomic eigenstates are non-trivial 
linear combinations of one-particle product states,
the hopping reduction factors 
are arranged in a $2N\times 2N$~matrix 
$q_{\bbox{\sigma}}^{\bbox{\sigma'}}$ with spin-orbit 
indices $\bbox{\sigma}$, $\bbox{\sigma'}$. These quantities
are non-trivial functions of (i)~the variational parameters $m_{i;\Gamma}$, 
(ii)~the local occupancies of the Hartree--Fock wave function, and (iii)~the 
one-electron densities in the interacting case. The derivation
of the latter requires the diagonalization of a $2N\times 2N$~matrix 
$\tensor{Z}$. Further complexities occur for the most general case, i.e.,
for an extended spin-orbit basis with more than one orbital-type per 
representation of the symmetry group of the atomic site. 
Here we have nonzero values for the orbital-nondiagonal parts of the
on-site one-particle expectation values. At the expense of another
unitary transformation this most general case is also covered by
our formalism. Naturally, this leads to a more complicated
structure for the matrices $\tensor{q}$ and $\tensor{Z}$
and the effective local hybridizations and one-particle densities.

Like the density-functional theory the variational method is intrinsically
limited to the description of ground-state properties, e.g., the ground-state
energy, compressibility, magnetization, and magnetic susceptibility. Similar
to the density-functional theory our variational approach naturally extends to
finite temperatures and low-frequency excitations since our variational
ground-state energy corresponds to that of an effective one-particle
Hamiltonian. The ``correlated bands'' of this Hamiltonian can be used for a
comparison with measured dispersion curves and effective masses. Note that
our approach is completely general and applies to all multi-band systems.
Therefore, we hope that it will be fruitful for the description of
correlated electron systems in the metallic phase. Naturally, any
quasi-particle approach is limited to the region of the validity of
Fermi-liquid theory.

In this work we presented explicit results for a degenerate two-band model
as the simplest non-trivial application of our method. We assumed $e_g$-type
orbitals on sites of a simple cubic lattice. For the single-particle 
Hamiltonian nearest and next-nearest neighbor transfer matrix elements
were used which give rise to two bands of width~$W=6.6\,$eV.
In our model we included all possible two-particle interactions. 
Yet, there exist only two independent interaction parameters~$U$,~$J$ 
since the relation $U-U'=2J$ holds due to symmetry, 
and, likewise, $J_{\text{C}}=J$ is fulfilled for the charge
exchange term. For our simple model system the $\tensor{Z}$~matrix 
is the unit matrix and the hopping reduction matrix is diagonal.

As a first application we studied the Brinkman--Rice metal--insulator
transition at half band-filling. Above some finite interaction strength all
electrons localize. As for the one-band case this localization transition is
rather questionable as a scenario for the Mott transition in multi-band
Mott--Hubbard systems. The lattice sites will not be isolated as in the
Brinkman--Rice insulator but they will remain coupled via the itinerant
exchange. Thus, in the large-$U$ limit, we should expect an antiferromagnetic 
insulating ground state for $J>0$ whereas, for $J<0$, antiferro-type orbital
ordering appears to be most likely.

Apart from the singular point~$J=0$ the metal-insulator transition 
is discontinuous as manifests itself in finite changes~$q_{\text{BR}}$
of the bandwidth reduction factor at the transition.
The results for Gutzwiller wave functions with atomic correlations
significantly differ from those for pure density correlations.
In particular, this applies to the behavior
of the curves~$q_{\text{BR}}(J/U_{\text{BR}})$ for large~$J$
which monotonously increases (decreases) as a function of~$J$ 
for atomic (pure density) correlations for $J/U>0.05$.
The Gutzwiller wave function with atomic correlations
is seen to be more ``flexible'' than that with pure density correlations
in the sense that the metallic state can much better adapt itself
to the insulator. 

The general aspect of a discontinuous
metal--insulator transition could be generic for multi-band Hubbard models. 
In the insulator the atoms are
dominantly in a specific ${\frak n}$-electron ground state 
which is compatible with Hund's first rule.
Other ${\frak n}$-electron states (of excitation energy~$J$),
even more ${\frak n}\pm 1$ electron states (excitation energies~$U$),
are separated from the ground state by finite gaps.
In the metallic phase a macroscopic number of energetically unfavorable
${\frak n}\pm 1$ electron states is created and, consequently,
also a macroscopic number of the other ${\frak n}$-electron states.
It appears to be unlikely that all of these occupation densities 
change continuously at a {\em single\/} critical interaction strength.
Instead, the metallic state breaks down discontinuously when
the gain in kinetic energy can no longer compensate the intra-atomic gaps. 
However, variational statements on the nature of the transition
between the metal and the Mott--Hubbard insulator must be taken with 
great care.~\cite{Mottbuch,BUCH} 

Nonetheless, for reasonably small~$J/U$ and 
for the case of a general band structure without the
perfect-nesting property we expect a transition to an
antiferromagnetic state with strong electronic localization, i.e.,
with large charge-transfer excitation energies. 
We do not expect transitions to an antiferromagnetic metallic 
or small-gap insulating state. Yet to test 
this conjecture an antiferromagnetic trial state needs to be investigated.

As a second application we addressed the issue of itinerant ferromagnetism. 
For this purpose we chose two band fillings, the first one
at the maximum of the density of states, the second one close to it.
Again, we find a large ``flexibility'' of the Gutzwiller wave function
with atomic correlations: the paramagnetic state accommodates
large local spins, as much as 90\% of the saturation value, and 
only a small jump is observed at the ferromagnetic transition.
Hence, the paramagnetic metallic state near the transition and, moreover, 
the ferromagnetic state are highly correlated.

In general, the ferromagnetic transition is found to occur
at fairly large interaction 
strengths~$U_{\text{F}}$, with values $1<U_{\text{F}}/W \lesssim 2$.
In addition, finite values of the exchange interaction~$J$ are required with
$J_{\text{min}}\approx 0.1U$. These results may change towards smaller values
of $(J,U)$ if a larger value for the peak density of states is chosen.
In any case, our results stress the importance of the atomic 
Hund's-rule exchange for ferromagnetism in multi-band models, a view fostered
a long time ago by van Vleck.~\cite{vanVleck}
The ferromagnetic condensation energy is an estimate for
$k_{\text{B}}T_{\text{C}}$, the Curie temperature for iron-group metals.
It is found to be of right order of magnitude, $T_{\text{C}}\approx
500\, \text{K}$, for interaction strengths as large as $10\, \text{eV}$. 
The condensation energy is a smooth function of the interaction strength, 
i.e., the ground-state magnetization does not too sensitively depend on~$U$.

In contrast, the corresponding Hartree--Fock treatment yields completely
different results. The ferromagnetic transition
is predicted to occur for small values of~$U$,
$U_{\text{F}}^{\text{HF}}/W<1$. The spin exchange~$J$ 
is fairly unimportant and the magnetization saturates almost immediately
as a function of~$U$. Finally, apart from a small interval 
above~$U_{\text{F}}^{\text{HF}}$, the condensation energy is grossly 
overestimated. 
Thus, itinerant ferromagnetism in interacting
multi-band systems is a correlated-electron problem that cannot be treated
within a weak-coupling approach.

The inferior results of the Hartree--Fock treatment
might be taken as an indication that spin-density functional theory is 
also inadequate for the description of itinerant ferromagnetism in iron-group 
metals because our results suggest strong correlation effects there.
On the other hand, the success of this effective single-particle theory
may point to an inadequacy of our multi-band model for the following reason. 
The results from spin-density functional theory indicate that
the minority spin bands are broader and, accordingly, the corresponding
wave functions more extended in space than those of the majority bands.
This ``orbital flexibility'' makes the minority spin density to
dominate in the interstitial regions. Orbital
flexibility is not included in the present form of our multi-band models. If
considered, e.g., by extending the orbital basis, the required interaction
strength for ferromagnetism may be reduced considerably towards a less
correlated situation. In principle, our treatment of generalized Gutzwiller
wave functions allows us to incorporate such basis extensions.
Work in this direction is in progress, and applications of our general 
formalism to real systems are presently under investigation.

\acknowledgments

J.~B.\ thanks P.~Nozi\`eres for an invitation to the ILL where this project
was started.

\appendix
\section*{Evaluation of expectation values}

In the appendix we sketch the essential steps for the exact evaluation of
Gutzwiller-correlated wave functions in infinite dimensions. First, we
choose a new basis in which local Fock terms are absent. Second, we select
appropriate expansion parameters for a perturbation theory around the limit
of zero interactions. As a third step we set up a diagrammatic theory for
the calculation of expectation values based on Wick's theorem and the linked
cluster theorem.~\cite{FetterWalecka}
The clue to the exact solution in infinite dimensions is the selection of
the expansion parameters. They are chosen in such a way that all higher
order diagrams vanish in infinite dimensions, and the trivial order gives
the exact result. Further technical details can be found in 
Refs.~\onlinecite{GebhardPRB1,GebhardPRB2,BGW}.
In the rest of the appendix we derive explicit results for the
local multiple occupancies and the interacting one-particle density matrix.

\subsection{Change of the local basis}
\label{basischangelocal}

For a general~$|\Phi_0\rangle $ the non-interacting
local one-particle matrix $\tensor{C}_i^0$ with the entries 
\begin{equation}
C_{i;\bbox{\gamma},\bbox{\gamma'}}^0
=\langle \Phi_0|\hat{c}_{i;\bbox{\gamma}}^{+}
\hat{c}_{i;\bbox{\gamma'}}^{}|\Phi_0\rangle
\end{equation}
is not diagonal. Therefore, we derive the formalism here which covers the
most general case.

Since our diagrammatic approach for the evaluation of Gutzwiller-correlated
wave functions requires that local Fock terms are absent, we need to perform
a local unitary transformation, 
\begin{mathletters}
\label{hh7}
\begin{eqnarray}
\sum_{\bbox{\gamma}}F_{i;\bbox{\sigma},\bbox{\gamma}}^{+}
F_{i;\bbox{\gamma},\bbox{\sigma'}}^{\vphantom{+}} &=&
\delta_{\bbox{\sigma},\bbox{\sigma'}}\;,
\\[3pt]
\hat{h}_{i;\bbox{\sigma}}^{+}=
\sum_{\bbox{\gamma}}F_{i;\bbox{\sigma},\bbox{\gamma}}^{+}
\hat{c}_{i;\bbox{\gamma}}^{+}\; &,&\;
\hat{c}_{i;\bbox{\gamma}}^{+}=
\sum_{\bbox{\sigma}}F_{i;\bbox{\gamma},\bbox{\sigma}}^{\vphantom{+}}
\hat{h}_{i;\bbox{\sigma}}^{+}\;, \\[3pt]
\hat{h}_{i;\bbox{\sigma}}^{\vphantom{+}}=
\sum_{\bbox{\gamma}}F_{i;\bbox{\gamma},\bbox{\sigma}}^{\vphantom{+}}
\hat{c}_{i;\bbox{\gamma}}^{\vphantom{+}}\; &,&\;
\hat{c}_{i;\bbox{\gamma}}^{\vphantom{+}}=
\sum_{\bbox{\sigma}}F_{i;\bbox{\sigma},\bbox{\gamma}}^+
\hat{h}_{i;\bbox{\sigma}}^{\vphantom{+}}\;.
\end{eqnarray}\end{mathletters}% HERE!
It diagonalizes the non-interacting local one-particle density matrix 
\begin{mathletters}
\label{hh8}
\begin{equation}
\left( \tensor{F}_i\right) ^{+}\tensor{C}_i^0 \tensor{F}_i=
\text{diag}(n_{i;\bbox{\sigma}}^{h,0})\;.  \label{matrixdefH}
\end{equation}
This is always possible because $\tensor{C}_i^0$ is Hermitian. 
As seen from~(\ref{matrixdefH}), local Fock terms are absent in the new basis, 
i.e., the non-interacting local one-particle
density matrix $\tensor{H}_i^0$ is diagonal,
\begin{equation}
H_{i;\bbox{\sigma},\bbox{\sigma'}}^0 =
\langle \Phi_0|
\hat{h}_{i;\bbox{\sigma}}^{+}\hat{h}_{i;\bbox{\sigma'}}^{\vphantom{+}}
|\Phi_0\rangle =\delta_{\bbox{\sigma},\bbox{\sigma'}}\langle
\Phi_0|
\hat{h}_{i;\bbox{\sigma}}^{+}\hat{h}_{i;\bbox{\sigma}}^{\vphantom{+}}
|\Phi_0\rangle =\delta_{\bbox{\sigma},\bbox{\sigma'}}n_{i;\bbox{\sigma}}^{h,0}
\;.  \label{noFOCK}
\end{equation}\end{mathletters}% HERE!
For a given $|\Phi_0\rangle $ the transformation matrix $\tensor{F}_i$ is
fixed, and the local occupancies $n_{i;\bbox{\sigma}}^{h,0}$ in the new
basis are the eigenvalues of $\tensor{C}_i^0$.

{}From now on we work in the new local basis. We suppress the site index for
the rest of this subsection. The notation of Sect.~\ref{atomicDEF} remains
essentially the same but each operator $\hat{c}_{\bbox{\sigma}}^+$ 
($\hat{c}_{\bbox{\sigma}}^{\vphantom{+}}$) has to be replaced by 
$\hat{h}_{\bbox{\sigma}}^+$ ($\hat{h}_{\bbox{\sigma}}^{\vphantom{+}}$). 
In the new basis the configuration eigenstates are denoted by 
\begin{mathletters}
\label{hh9}
\begin{equation}
|{\cal H}\rangle = \prod_{n=1}^{|{\cal H}|} \hat{h}_{\bbox{\sigma_n}}^+ 
|\text{vacuum}\rangle \quad (\bbox{\sigma_n}\in {\cal H}) \; ,
\end{equation}
and the atomic eigenstates~$|\Gamma\rangle$ and their projection operator 
$m_{\Gamma}=|\Gamma\rangle\langle \Gamma|$ are given by 
\begin{eqnarray}
|\Gamma \rangle &=& \sum_{{\cal H}} A_{{\cal H},\Gamma} |{\cal H}\rangle \;,
\\[3pt]
\hat{m}_{\Gamma} &=& \sum_{{\cal H}, {\cal H}'} A_{{\cal H},\Gamma} 
\hat{m}_{{\cal H},{\cal H}'} A_{\Gamma,{\cal H}'}^+ \; .
\label{mGammaAmatrix}
\end{eqnarray}\end{mathletters}% HERE!
The elements of the unitary matrix $\tensor{A}$ are given by 
\begin{mathletters}
\label{defAandus}
\begin{eqnarray}
A_{{\cal H},\Gamma} &=& \langle {\cal H} | \Gamma\rangle = \sum_I
T_{I,\Gamma} \langle {\cal H} | I \rangle \; ,  \nonumber \\[3pt]
\langle {\cal H} | I \rangle &=& 
\text{det} (F_{\bbox{\gamma_i},\bbox{\sigma_j}}) 
\quad , \quad (\bbox{\gamma_i} \in I, \bbox{\sigma_j} \in {\cal H}) \; .
\label{defAbyu}
\end{eqnarray}
The inverse relation to~(\ref{defAbyu}) reads 
\begin{equation}
T_{I,\Gamma} = \sum_{{\cal H}} A_{{\cal H},\Gamma} 
\langle I | {\cal H} \rangle \;.  \label{defubyA}
\end{equation}\end{mathletters}% HERE!
Again, $\tensor{A}$ is block-diagonal. Eqs.~(\ref{defAandus}) are defined
for $|\Gamma|=|{\cal H}|\geq 2$. For $\Gamma={\cal H}=\emptyset$ we set
$A_{\emptyset,\emptyset}=1$. 
The entries in $\tensor{A}$ for $|\Gamma|=|{\cal H}|=1$ can be chosen 
at our convenience. We will specify them such that an
exact evaluation of our variational wave functions becomes feasible in
infinite dimensions; see below.

After the change of the basis we obtain simple expressions for expectation
values in $|\Phi_0\rangle$ with the help of Wick's 
theorem.~\cite{FetterWalecka} For example, we have 
\begin{eqnarray}
m_{\Gamma}^0 &=& \sum_{{\cal H}} |A_{{\cal H},\Gamma}|^2 m_{{\cal H}}^{h,0}
\;,  \nonumber \\[3pt]
m_{{\cal H}}^{h,0} &=& \prod_{\bbox{\sigma}\in {\cal H}} n_{\bbox{\sigma}}^{h,0}
\prod_{\bbox{\sigma}\in \overline{{\cal H}}} 
\Bigl(1-n_{\bbox{\sigma}}^{h,0} \Bigr) \;,
\end{eqnarray}
where we used the fact 
that Fock terms are absent in the new basis; see eq.~(\ref{noFOCK}).

\subsection{Choice of the expansion parameter}

\label{choice}

In this subsection we suppress the site index. We proceed along the
derivation outlined in Refs.~\onlinecite{GebhardPRB1,GebhardPRB2,BGW}.
First, we express the square of the Gutzwiller 
correlator~$\hat{P}_{\text{G}}^2$ in terms of the 
operators for the configuration eigenstates, 
\begin{eqnarray}
\hat{P}_{\text{G}}^2 
&=& 
1+ \sum_{\Gamma} (\lambda_{\Gamma}^2-1) \hat{m}_{\Gamma} =
1+\sum_{{\cal H},{\cal H}'} y_{{\cal H},{\cal H}'} \hat{m}_{{\cal H},{\cal H}'}
\; ,  \nonumber \\[3pt]
y_{{\cal H},{\cal H}'} &=& \sum_{\Gamma} (\lambda_{\Gamma}^2-1) 
A_{{\cal H},\Gamma} A_{\Gamma,{\cal H}'}^+ \; .  \label{defyvalues}
\end{eqnarray}
Next, we demand that 
\begin{equation}
\sum_{{\cal H},{\cal H}'} y_{{\cal H},{\cal H}'} 
\hat{m}_{{\cal H},{\cal H}'} 
= \sum_{{\cal H},{\cal H}'\, (|{\cal H}|,|{\cal H}'|\geq 2)} 
x_{{\cal H},{\cal H}'} \hat{n}_{{\cal H},{\cal H}'}^{\text{HF}} 
\; ,  \label{countvariables}
\end{equation}
such that 
\begin{equation}
\hat{P}_{\text{G}}^2 
= 1+ \sum_{{\cal H},{\cal H}'\, (|{\cal H}|,|{\cal H}'|\geq 2)} 
x_{{\cal H},{\cal H}'} \hat{n}_{{\cal H},{\cal H}'}^{\text{HF}} 
\; .  \label{defxvalues}
\end{equation}
Here, 
\begin{mathletters}
\label{HFopDEF}
\begin{eqnarray}
\hat{n}_{{\cal H},{\cal H}}^{\text{HF}}&= &\hat{n}_{{\cal H}}^{\text{HF}}=
\prod_{\bbox{\sigma}\in {\cal H}} \hat{n}_{\bbox{\sigma}}^{\text{HF}} \; , 
\\[3pt]
\hat{n}_{\bbox{\sigma}}^{\text{HF}}&=& 
\hat{n}_{\bbox{\sigma}}^h -n_{\bbox{\sigma}}^{h,0}
\end{eqnarray}
for ${\cal H}={\cal H}'$, and 
\begin{equation}
\hat{n}_{{\cal H},{\cal H}'}^{\text{HF}}
= \biggl[ \prod_{\bbox{\sigma}\in {\cal J}} 
\hat{n}_{\bbox{\sigma}}^{\text{HF} }\biggr] \hat{n}_{{\cal H}_1,{\cal H}_2} 
\qquad 
({\cal J}={\cal H}\cap {\cal H}';
{\cal H}={\cal J}\cup {\cal H}_1;{\cal H}'={\cal J}\cup {\cal H}_2)
\end{equation}\end{mathletters}% HERE!
for ${\cal H}\neq{\cal H}'$; compare Sects.~\ref{atomicDEF} 
and~\ref{basischangelocal}. 
Note that $\langle \Phi_0 | \hat{n}_{{\cal H}}^{\text{HF}} 
| \Phi_0 \rangle=0$ because we subtracted the Hartree terms, and all Fock
terms vanish in~$| \Phi_0 \rangle$ due to eq.~(\ref{noFOCK}).

The expansion of~$\hat{P}_{\text{G}}^2$ in~(\ref{defxvalues}) is chosen such
that at least four lines meet at every internal vertex in our diagrammatic
expansion; see below. The number of 
parameters~$x_{{\cal H},{\cal H}'}$ in~(\ref{countvariables}) is 
less than the number of parameters~$y_{{\cal H},{\cal H}'}$ 
due to the restriction $|{\cal H}|=|{\cal H}'|\geq 2$, i.e., 
we essentially require 
\begin{mathletters}
\label{HH10}
\begin{eqnarray}
x_{\emptyset,\emptyset}&=& 0 \; ,  \label{RHSanew} \\[3pt]
x_{\bbox{\sigma},\bbox{\sigma'}} &=& 0 \; .  \label{RHSbnew}
\end{eqnarray}\end{mathletters}% HERE!
Alternatively, as follows from~(\ref{defxvalues}), these $1+(2N)^2$ local
conditions can be formulated as 
\begin{mathletters}
\label{RHS}
\begin{eqnarray}
\langle \Phi_0 | \hat{P}_{\text{G}}^2 | \Phi_0\rangle &=& 1\; ,  \label{RHSa}
\\[3pt]
\langle \Phi_0 | 
\hat{h}_{\bbox{\sigma}}^+ \hat{h}_{\bbox{\sigma'}}^{\vphantom{+}} 
\hat{P}_{\text{G}}^2 | \Phi_0\rangle 
&=& \langle \Phi_0 | 
\hat{h}_{\bbox{\sigma}}^+ \hat{h}_{\bbox{\sigma'}}^{\vphantom{+}} 
| \Phi_0\rangle
\; .  \label{RHSb}
\end{eqnarray}\end{mathletters}% HERE!
The first equation follows immediately because we eliminated all Hartree
terms from the right-hand-side of~(\ref{defxvalues}). For the other $(2N)^2$
equations~(\ref{RHSb}) we analyze the case $\bbox{\sigma}=\bbox{\sigma'}$
first. The operators $\hat{n}_{{\cal H}}^{\text{HF}}$ on the right-hand-side
of~(\ref{defxvalues}) contain at least two Hartree--Fock operators 
$(\hat{n}_{\bbox{\sigma_1}}^h-n_{\bbox{\sigma_1}}^{h,0}) 
(\hat{n}_{\bbox{\sigma_2}}^h-n_{\bbox{\sigma_2}}^{h,0})$ 
which cannot be eliminated completely by a
single operator $\hat{n}_{\bbox{\sigma}}^h$. 
The term with $\hat{n}_{{\cal H},{\cal H}'}^{\text{HF}}$ 
for ${\cal H}\neq {\cal H}'$ vanishes because of the Fock terms 
which transfer electrons from~${\cal H}'$ to~${\cal H}$. 
According to eq.~(\ref{noFOCK}) the expectation
value of Fock terms vanishes in~$|\Phi_0\rangle$. 
For $\bbox{\sigma}\neq \bbox{\sigma'}$ we note that eq.~(\ref{HFopDEF}) 
requires ${\cal H'}=\bbox{\gamma}\cup \bbox{\sigma}$ 
and ${\cal H}=\bbox{\gamma}\cup \bbox{\sigma'}$ to eliminate all possible 
Fock terms. Nevertheless, this contribution still vanishes because of the 
remaining Hartree--Fock operator $\hat{n}_{\bbox{\gamma}}^{\text{HF}}$ 
for the orbital~$\bbox{\gamma}$.

In Sect.~\ref{solutioneqA12} we shall give the explicit solution of the
equations~(\ref{RHS}) in infinite dimensions.

\subsection{Diagrammatic theory and simplifications in infinite dimensions}
\label{simplifyinfty}

Since the variational parameters obey $\lambda_{i;\Gamma}= 1$ in the absence
of interactions, the parameters~$x_{i;{\cal H},{\cal H}'}$ go to
zero for vanishing interactions. Therefore, they are suitable for a
perturbation expansion in which the order of the expansion is given by the
number of factors $x$. When we perform this expansion we may apply Wick's
theorem for the resulting expectation values since $|\Phi_0\rangle$ is a
one-particle state. As usual in perturbation theory around a single-particle
state~\cite{FetterWalecka} the resulting contributions can be represented
diagrammatically. In our theory the ``internal vertices'' represent the
factors~$x_{i;{\cal H},{\cal H}'}$. Besides these internal vertices
there are also ``external vertices'' which come from the site dependence of
the operators $\hat{O}$. For example, there are two external vertices at the
sites~$i$ and $j$ for 
$\hat{O}= \hat{h}_{i;\bbox{\sigma}}^+\hat{h}_{j;\bbox{\sigma'}}^{\vphantom{+}}$.
The non-trivial result in infinite
dimensions stem from the Hartree contributions at the external vertices;
see below.

To obtain an expansion in powers of~$x$ we set $\hat{P}_{i;\text{G}}^2 =1+ 
\overline{P}_i$ and write 
\begin{equation}
\prod_i \left[ 1 + \overline{P}_i\right] = 
1 + \sum_{k=1}^{\infty} \frac{1}{k!} \mathop{{\sum}'}_{i_1,\ldots,i_k} 
\prod_{j=i_1}^{i_k} \overline{P}_j \; ,
\end{equation}
where $i_1,\ldots,i_k$ specify internal vertices. Here, the prime on the sum
indicates that all lattice sites $i_1, \ldots, i_k$ are different when we
apply Wick's theorem. Consequently, the ``lines'' of our diagrammatic theory
are given by the one-particle density matrix for the single-particle wave
function $|\Phi_0\rangle$ for $i\neq j$, 
\begin{equation}
P_{i,j}^{\bbox{\sigma},\bbox{\sigma'}}=(1-\delta_{i,j}) \langle \Phi_0 | 
\hat{h}_{i;\bbox{\sigma}}^{+} \hat{h}_{j;\bbox{\sigma'}}^{\vphantom{+}} 
|\Phi_0\rangle \equiv (1-\delta_{i,j}) \langle \hat{h}_{i;\bbox{\sigma}}^{+} 
\hat{h}_{j;\bbox{\sigma'}}^{\vphantom{+}} \rangle_0 \; .  \label{20}
\end{equation}
Note that we do not have to distinguish between ``hole'' and ``particle''
lines because all sites are different when we apply Wick's 
theorem.~\cite{GebhardPRB1,GebhardPRB2,BGW}

To make further progress we have to apply the linked cluster 
theorem.~\cite{FetterWalecka} 
Unfortunately, the restriction on the lattice sums prevents
its direct application. As shown in Ref.~\onlinecite{GebhardPRB2,BGW} this
problem can be circumvented by a redefinition of the internal vertices,
i.e., $x_{i;{\cal H},{\cal H}'} \to \widetilde{x}_{i;{\cal H},{\cal H}'}$. 
As a result we obtain a standard diagrammatic theory with
renormalized vertices $\widetilde{x}_{i;{\cal H},{\cal H}'}$ and
lines given by~(\ref{20}). Since the trivial order does not contain any
internal vertex, it is unaffected by the redefinition of the internal
vertices.

In our theory we subtracted the Hartree contributions and ruled out local
Fock terms according to~(\ref{noFOCK}). For our diagrams this implies that
there are no trivial loops at any internal vertex. Consequently, there are
at least three independent paths from one vertex~$i_1$ to another
vertex~$i_2 $ in each non-trivial diagram since $|{\cal H}|=|{\cal H}'|\geq 2 $
requires that at least four lines meet at every internal vertex; paths
are independent if they do not have a line in common. In the limit of
infinite dimensions~\cite{MVPRLdinfty} only $i_1=i_2$ contributes to a
diagram if $i_1 $ and $i_2$ are linked by (at least) three independent
paths. In our case this implies that the diagram simply {\em vanishes\/}
since a line linking two identical vertices is zero by the 
definition~(\ref{20}).~\cite{GebhardPRB1,GebhardPRB2,BGW} 
Consequently, the trivial order of
our expansion gives the exact result in infinite dimension, e.g., only the
diagram for 
$\langle \hat{h}_{i;\bbox{\sigma}}^+ \hat{h}_{j;\bbox{\sigma'}}^{\vphantom{+}}
\rangle$ with a single line survives the limit $d\to\infty$.
The remaining task is the calculation of the trivial (Hartree) terms which
stem from the external vertices. This will be carried out for the local
occupancies and the one-particle density matrix in the rest of the appendix.

\subsection{Local atomic occupancies}
\label{atomicocc}

In this subsection we suppress the site index. We need to evaluate 
$\langle m_{\Gamma} \rangle$. 
As described in the previous subsection, the task is
readily solved in infinite dimension, 
\begin{equation}
m_{\Gamma} =\langle \hat{m}_{\Gamma} \rangle 
=\langle \hat{P}_{\text{G}}\hat{m}_{\Gamma} 
\hat{P}_{\text{G}}\rangle_0=\lambda_{\Gamma}^2m_{\Gamma}^0\;,
\label{localoccappendix}
\end{equation}
where we used the definitions~(\ref{mgamma}) and~(\ref{GutzcorrdegbandsHund}). 
This proves 
eq.~(\ref{HundgutzrelationsA}). 
Furthermore, we may use this result to show that
the condition~(\ref{RHSa}) is indeed fulfilled. We have 
\begin{equation}
\langle \Phi_0|\hat{P}_{\text{G}}^2|\Phi_0\rangle =\langle \sum_{\Gamma}
\lambda_{\Gamma}^2\hat{m}_{\Gamma} \rangle_0=\sum_{\Gamma} m_{\Gamma} =1\;,
\end{equation}
because the local completeness relation~(\ref{localcompelete}) holds for the
correlated wave function in any dimension.

For later use we write $\hat{P}_{\text{G}}$ in the form 
\begin{mathletters}
\label{splitPGlambda}
\begin{equation}
\hat{P}_{\text{G}}=\sum_{{\cal H},{\cal H}'}\lambda_{{\cal H},{\cal H}'}
\hat{m}_{{\cal H},{\cal H}'}\;,
\label{PfunctII'}
\end{equation}
where we defined 
\begin{equation}
\lambda_{{\cal H},{\cal H}'}=\sum_{\Gamma} A_{{\cal H},\Gamma}
\lambda_{\Gamma}A_{\Gamma ,{\cal H}'}^{+}\;.  \label{deflambdaiII'}
\end{equation}\end{mathletters}% HERE!
In infinite dimensions we then have 
\begin{mathletters}
\label{HundgutzrelationsCappendixglobal} 
\begin{eqnarray}
m_{{\cal H},{\cal H}'} 
&=&
\langle \hat{m}_{{\cal H},{\cal H}'}\rangle 
=\langle \hat{P}_{\text{G}}\hat{m}_{{\cal H},{\cal H}'}
\hat{P}_{\text{G}}\rangle_0  \nonumber \\[3pt]
&=&\sum_{{\cal H}_1,{\cal H}_2,{\cal H}_3,{\cal H}_4}
\lambda_{{\cal H}_1,{\cal H}_2}
\lambda_{{\cal H}_3,{\cal H}_4}
\langle \hat{m}_{{\cal H}_1,{\cal H}_2}
\hat{m}_{{\cal H},{\cal H}'}\hat{m}_{{\cal H}_3,{\cal H}_4}
\rangle_0  \label{HundgutzrelationsCappendixA} \\[3pt]
&=&\sum_{{\cal K}}\lambda_{{\cal K},{\cal H}}\lambda_{{\cal H}',{\cal K}}
m_{{\cal K}}^{h,0} \; , \nonumber
\end{eqnarray}
where we used~(\ref{projectoridentity}) in the last step. 
In particular, for ${\cal H}={\cal H'}$ we obtain 
\begin{equation}
m_{{\cal H}} =  m_{{\cal H},{\cal H}} =  
\sum_{\cal K} \bigl| \lambda_{{\cal K},{\cal H}}\bigr|^2 m_{{\cal K}}^{h,0}
\label{HundgutzrelationsCappendix} 
\end{equation}\end{mathletters}%
With the help of~(\ref{localoccappendix}) and the 
definition~(\ref{deflambdaiII'}) eq.~(\ref{mImGammastrich}) follows.

Now we solve the $(2N)^2$ 
equations~(\ref{RHSb}). We multiply both sides of~(\ref{defxvalues}) 
with $\hat{h}_{\bbox{\sigma}}^+ \hat{h}_{\bbox{\sigma'}}^{\vphantom{+}}$ 
and take the expectation value
with respect to~$|\Phi_0\rangle$. With the help of eqs.~(\ref{localcompelete}) 
and~(\ref{mGammaAmatrix}) eq.~(\ref{RHSb}) becomes 
\begin{equation}
\langle \hat{h}_{\bbox{\sigma}}^+ \hat{h}_{\bbox{\sigma'}}^{\vphantom{+}}
\rangle_0 = \sum_{\Gamma} \lambda_{\Gamma}^2 
\sum_{{\cal H},{\cal H}'} A_{{\cal H},\Gamma}A_{\Gamma,{\cal H}'}^+ 
\langle \hat{h}_{\bbox{\sigma}}^+ \hat{h}_{\bbox{\sigma'}}^{\vphantom{+}} 
\hat{m}_{{\cal H},{\cal H}'} \rangle_0 \; .  \label{Weberseq}
\end{equation}
Apparently we may set ${\cal H}={\cal J}\cup \bbox{\sigma'}$ ($\bbox{\sigma'}
\not\in {\cal J}$). With the help of~(\ref{noFOCK}) we then find 
that ${\cal H}'= {\cal J}\cup \bbox{\sigma}$ and, therefore, 
$\bbox{\sigma}\not\in {\cal J}$. 
We use the definition of the fermionic sign function~(\ref{deffsign}) 
and~(\ref{localoccappendix}) which allow us to simplify the
above equation to 
\begin{mathletters}
\label{DEFZ}
\begin{equation}
Z_{\bbox{\sigma},\bbox{\sigma'}} = \sum_{|\Gamma|=1} 
A_{\bbox{\sigma},\Gamma}^* \lambda_{\Gamma}^2 
\Bigl(A_{\Gamma,\bbox{\sigma'}}^+\Bigr)^* \; ,
\label{DEFZa}
\end{equation}
where the entries of the $(2N) \times (2N)$~matrix $\tensor{Z}$ are given by 
\begin{equation}
Z_{\bbox{\sigma},\bbox{\sigma'}} 
= \frac{n_{\bbox{\sigma}}^{h,0}}{m_{\bbox{\sigma}}^{h,0}} 
\delta_{\bbox{\sigma},\bbox{\sigma'}} -
\sum_{|\Gamma|\geq 2} \frac{m_{\Gamma}}{m_{\Gamma}^0} 
\sum_{{\cal J}\, (\bbox{\sigma},\bbox{\sigma'} \not\in {\cal J})} 
\text{fsgn}(\bbox{\sigma'},{\cal J}) 
\text{fsgn}(\bbox{\sigma},{\cal J}) 
A_{({\cal J}\cup\bbox{\sigma'}),\Gamma} 
A_{\Gamma,({\cal J}\cup\bbox{\sigma})}^+ 
\frac{m_{{\cal J}\cup(\bbox{\sigma},\bbox{\sigma'})}^{h,0} }%
{m_{(\bbox{\sigma},\bbox{\sigma'})}^{h,0}} \; .  \label{defASIGMAsigma}
\end{equation}\end{mathletters}% HERE!
We can safely ignore the unphysical case of $m_{\bbox{\sigma}}^{h,0}=0$. All
quantities in the matrix $\tensor{Z}$ are known as soon as we fix 
$|\Phi_0\rangle$ and our variational parameters~$m_{\Gamma}$ for 
$|\Gamma|\geq 2$. Eq.~(\ref{DEFZa}) states that a unitary 
$(2N)\times (2N)$-matrix $\tensor{A}'$
with $A_{\bbox{\sigma},\Gamma}'\equiv A_{\bbox{\sigma},\Gamma}^*$
diagonalizes the Hermitian
matrix~$\tensor{Z}$, and that $\lambda_{\Gamma}^2\geq 0$ ($|\Gamma|=1$) are
its eigenvalues, 
\begin{equation}
(\tensor{A}')^+ \tensor{Z} (\tensor{A}') = 
\text{diag}(\lambda_{\Gamma}^2) \; .
\end{equation}
Therefore, the conditions~(\ref{RHSb}) fix the matrix 
$A_{\bbox{\sigma},\Gamma}$ for $|\Gamma|=|\bbox{\sigma}|=1$, and the 
expectation values for the atomic
configurations with a single electron are given by
$m_{\Gamma}=\lambda_{\Gamma}^2 m_{\Gamma}^{0}$.

\subsection{$\bbox{\tensor{q}}$~matrix}
\label{kinenergy}

As in the previous subsection we have to work in the new basis. Therefore,
we start the derivation of the $\tensor{q}$~matrix with a unitary
transformation, 
\begin{equation}
\langle 
\hat{c}_{i;\bbox{\gamma_{1}}}^{+}
\hat{c}_{j;\bbox{\gamma_{1}'}}^{\vphantom{+}}
\rangle 
=\sum_{\bbox{\sigma_{1}},\bbox{\sigma_{1}'}}
F_{i;\bbox{\gamma_{1}},\bbox{\sigma_{1}}}
F_{j;\bbox{\sigma_{1}'},\bbox{\gamma_{1}'}}^{+}
\langle \hat{h}_{i;\bbox{\sigma_{1}}}^{+}
\hat{h}_{j;\bbox{\sigma_{1}'}}^{\vphantom{+}}\rangle \;.
\end{equation}
In Sect.~\ref{simplifyinfty} we showed that the calculation of the
interacting one-particle density matrix reduces to 
\begin{equation}
\langle 
\hat{h}_{i;\bbox{\sigma_{1}}}^{+}
\hat{h}_{j;\bbox{\sigma_{1}'}}^{\vphantom{+}}
\rangle =
\langle \Phi_0|
\Bigl(\hat{P}_{i;\text{G}}
\hat{h}_{i;\bbox{\sigma_{1}}}^{+}
\hat{P}_{i;\text{G}}\Bigr)\Bigl(
\hat{P}_{j;\text{G}}\hat{h}_{j;\bbox{\sigma_{1}'}}^{\vphantom{+}}
\hat{P}_{j;\text{G}}\Bigr)
|\Phi_0\rangle  \label{localqinfty}
\end{equation}
in infinite dimensions. There we also showed that only a single line can
join the two external vertices~$i$ and~$j$. This implies 
\begin{mathletters}
\label{Wickapp}
\begin{eqnarray}
\langle 
\hat{h}_{i;\bbox{\sigma_{1}}}^{+}
\hat{h}_{j;\bbox{\sigma_{1}'}}^{\vphantom{+}}
\rangle 
&=&
\sum_{\bbox{\sigma_{2}},\bbox{\sigma_{2}'}}
\sqrt{q_{i;\bbox{\sigma_{1}}}^{\bbox{\sigma_{2}}}
q_{j;\bbox{\sigma_{1}'}}^{\bbox{\sigma_{2}'}}}
\langle \Phi_0|
\hat{h}_{i;\bbox{\sigma_{2}}}^{+}
\hat{h}_{j;\bbox{\sigma_{2}'}}^{\vphantom{+}}
|\Phi_0\rangle  \label{Wickappa} \\[3pt]
&=&
\sum_{\bbox{\gamma_{2}},\bbox{\gamma_{2}'}}
\langle \hat{c}_{i;\bbox{\gamma_{2}}}^{+}
\hat{c}_{j;\bbox{\gamma_{2}'}}^{\vphantom{+}}\rangle_0
\sum_{\bbox{\sigma_{2}},\bbox{\sigma_{2}'}}
\sqrt{q_{i;\bbox{\sigma_{1}}}^{\bbox{\sigma_{2}}}
q_{j;\bbox{\sigma_{1}'}}^{\bbox{\sigma_{2}'}}}
F_{i;\bbox{\sigma_{2}},\bbox{\gamma_{2}}}^{+}
F_{j;\bbox{\gamma_{2}'},\bbox{\sigma_{2}'}}
\;,  \label{Wickappb}
\end{eqnarray}\end{mathletters}% HERE!
which proves the general structure of the variational kinetic 
energy~(\ref{widetildet}), 
\begin{mathletters}
\label{allresultsdegbandsHundappendix}
\begin{eqnarray}
\langle \hat{H}_1\rangle &=&
\sum_{i\neq j;\bbox{\gamma_{1}},\bbox{\gamma_{1}'}}
\widetilde{t}_{i,j}^{\,\bbox{\gamma_{1}},\bbox{\gamma_{1}'}}
\langle 
\hat{c}_{i;\bbox{\gamma_{1}}}^{+}
\hat{c}_{j;\bbox{\gamma_{1}'}}^{\vphantom{+}}
\rangle_0\;, \\[3pt]
\widetilde{t}_{i,j}^{\,\bbox{\gamma_{1}},\bbox{\gamma_{1}'}} 
&=&
\sum_{\bbox{\sigma_{1}},\bbox{\sigma_{1}'},\bbox{\sigma_{2}},\bbox{\sigma_{2}'}}
\sqrt{q_{i;\bbox{\sigma_{2}}}^{\bbox{\sigma_{1}}}
q_{j;\bbox{\sigma_{2}'}}^{\bbox{\sigma_{1}'}}}
F_{i;\bbox{\sigma_{1}},\bbox{\gamma_{1}}}^{+}
F_{j;\bbox{\gamma_{1}'},\bbox{\sigma_{1}'}}^{\vphantom{+}}
\sum_{\bbox{\gamma_{2}},\bbox{\gamma_{2}'}}
t_{i,j}^{\bbox{\gamma_{2}},\bbox{\gamma_{2}'}}
F_{i;\bbox{\gamma_{2}},\bbox{\sigma_{2}}}^{\vphantom{+}}
F_{j;\bbox{\sigma_{2}'},\bbox{\gamma_{2}'}}^{+}\;.  \label{widetildetappendix}
\end{eqnarray}\end{mathletters}% HERE!
Recall that the $\tensor{F}$~matrix is the unit matrix when
eq.~(\ref{noFOCKcops}) is fulfilled.

{}From now on we suppress the site index. 
To derive the explicit form of the 
$\tensor{q}$~matrix we use 
$\hat{P}_{\text{G}}$ in the form~(\ref{splitPGlambda})
to write 
\begin{equation}
\hat{P}_{\text{G}}
\hat{h}_{\bbox{\sigma}}^{+}\hat{P}_{\text{G}}=
\sum_{{\cal H}_1,{\cal H}_2,{\cal H}_3,{\cal H}_4}
\lambda_{{\cal H}_1,{\cal H}_2}\lambda_{{\cal H}_3,{\cal H}_4}
\hat{m}_{{\cal H}_1,{\cal H}_2}\hat{h}_{\bbox{\sigma}}^{+}
\hat{m}_{{\cal H}_3,{\cal H}_4}\;.
\end{equation}
We use the Dirac representation of the operators 
$\hat{m}_{{\cal H},{\cal H}'}=|{\cal H}\rangle \langle {\cal H}'|$ and find 
\begin{eqnarray}
\hat{P}_{\text{G}}\hat{h}_{\bbox{\sigma}}^{+}\hat{P}_{\text{G}} 
&=&
\sum_{{\cal H}_1,{\cal H}_2,{\cal H}_3,{\cal H}_4}
\lambda_{{\cal H}_1,{\cal H}_2}
\lambda_{{\cal H}_3,{\cal H}_4}
\langle {\cal H}_2|\hat{h}_{\bbox{\sigma}}^{+}|{\cal H}_3\rangle 
\hat{m}_{{\cal H}_1,{\cal H}_4}  \nonumber \\[3pt]
&=&
\sum_{{\cal H}_1,{\cal H}_2,{\cal H}_3,{\cal H}_4}
\lambda_{{\cal H}_1,{\cal H}_2}
\lambda_{{\cal H}_3,{\cal H}_4}
\text{fsgn}(\bbox{\sigma},{\cal H}_3)
\delta_{{\cal H}_2,{\cal H}_3\cup \bbox{\sigma}}
\hat{m}_{{\cal H}_1,{\cal H}_4}\;,
\end{eqnarray}
where we used the definition of the fermionic sign function~(\ref{deffsign}). 
Note that $\bbox{\sigma}\not\in {\cal H}_3$ is now required. We see that 
$|{\cal H}_1|=|{\cal H}_4|+1$, and our arguments presented in the last
subsection show that we have ${\cal H}_1={\cal H}'\cup \bbox{\sigma'}$ 
($\bbox{\sigma'}\not\in {\cal H}'$) and ${\cal H}_4={\cal H}'$ 
in infinite dimensions. Otherwise, local Fock terms
would appear in the evaluation of~(\ref{localqinfty}).
For $\bbox{\sigma'} \not\in {\cal H}'$ we introduce the operator 
\begin{equation}
\hat{m}_{{\cal H}',{\cal H}'}^{\bbox{\sigma'}}=
\prod_{\bbox{\gamma}\in {\cal H}'\backslash \bbox{\sigma'}}
\hat{n}_{\bbox{\gamma}}^h
\prod_{\bbox{\gamma}\in \overline{{\cal H}'}\backslash \bbox{\sigma'}}
(1-\hat{n}_{\bbox{\gamma}}^h)\;,
\end{equation}
which allows us to write (${\cal H}_3\equiv {\cal H}$) 
\begin{equation}
\hat{P}_{\text{G}}\hat{h}_{\bbox{\sigma}}^{+}\hat{P}_{\text{G}}=
\sum_{\bbox{\sigma'}} \hat{h}_{\bbox{\sigma'}}^{+}
\sum_{{\cal H}'\,(\bbox{\sigma'}\not\in {\cal H}')}
\sum_{{\cal H}\,(\bbox{\sigma} \not\in {\cal H})}
\lambda_{({\cal H}'\cup \bbox{\sigma'}),({\cal H}\cup \bbox{\sigma})}
\lambda_{{\cal H},{\cal H}'}
\text{fsgn}(\bbox{\sigma},{\cal H})
\text{fsgn}(\bbox{\sigma'},{\cal H}')
\hat{m}_{{\cal H}',{\cal H}'}^{\bbox{\sigma'}}  \label{Gndres}
\end{equation}
in infinite dimensions. When we compare this expression with~(\ref{Wickappa})
we see that 
\begin{equation}
\sqrt{q_{\bbox{\sigma}}^{\bbox{\sigma'}}}
=\sum_{{\cal H}'\,(\bbox{\sigma'}\not\in {\cal H}')}
\sum_{{\cal H}\,(\bbox{\sigma}\not\in {\cal H})}
\lambda_{({\cal H}'\cup \bbox{\sigma'}),({\cal H}\cup \bbox{\sigma})}
\lambda_{{\cal H},{\cal H}'}
\text{fsgn}(\bbox{\sigma},{\cal H})
\text{fsgn}(\bbox{\sigma'},{\cal H}')\langle 
\hat{m}_{{\cal H}',{\cal H}'}^{\bbox{\sigma'}}\rangle_0\;,
\end{equation}
because we finally singled out the electron creation 
operator~$\hat{h}_{\bbox{\sigma'}}^{+}$ for the contraction 
according to Wick's theorem in~(\ref{Wickappa}). 
We use eqs.~(\ref{localoccappendix}) and~(\ref{deflambdaiII'})
and the trivial relation 
\begin{equation}
\langle \hat{m}_{{\cal H}',{\cal H}'}^{\bbox{\sigma'}}\rangle_0
=\frac{m_{{\cal H}'}^{h,0}}{1-n_{\bbox{\sigma'}}^{h,0}}=
\sqrt{
\frac{m_{{\cal H}'}^{h,0}m_{({\cal H}'\cup \bbox{\sigma'})}^{h,0}}%
{n_{\bbox{\sigma'}}^{h,0}(1-n_{\bbox{\sigma'}}^{h,0})}}
\end{equation}
to derive the $\tensor{q}$~matrix in the form 
\begin{eqnarray}
\sqrt{q_{\bbox{\sigma}}^{\bbox{\sigma'}}} 
&=&
\sqrt{\frac{1}{n_{\bbox{\sigma'}}^{h,0}(1-n_{\bbox{\sigma'}}^{h,0})}}
\sum_{\Gamma,\Gamma'}
\sqrt{\frac{m_{\Gamma} m_{\Gamma'}}{m_{\Gamma}^0m_{\Gamma '}^0}}
\label{qmatrixappend} \\[3pt]
&&\times 
\sum_{
{{\ {\cal H},{\cal H}'} \atop 
{(\bbox{\sigma}\not\in {\cal H},\bbox{\sigma'}\not\in {\cal H}')}}}
\!\!\text{fsgn}(\bbox{\sigma'},{\cal H}')
\text{fsgn}(\bbox{\sigma},{\cal H})
\sqrt{m_{({\cal H}'\cup \bbox{\sigma'})}^{h,0}m_{{\cal H}'}^{h,0}}
A_{\Gamma,({\cal H}\cup \bbox{\sigma})}^{+}
A_{({\cal H}'\cup \bbox{\sigma'}),\Gamma }
A_{\Gamma',{\cal H}'}^{+}
A_{{\cal H},\Gamma'}\;.  \nonumber
\end{eqnarray}
When eq.~(\ref{noFOCKcops}) holds we recover eq.~(\ref{qmatrix}).

\subsection{Local one-particle density matrix}
\label{solutioneqA12}

Finally, we derive an expression for the {\em interacting\/} local
one-particle density matrix,
\begin{mathletters}
\begin{equation}
C_{\bbox{\gamma_1},\bbox{\gamma'_1}}
=\langle \hat{c}_{\bbox{\gamma_1}}^{+}
\hat{c}_{\bbox{\gamma'_1}}^{\vphantom{+}}
\rangle  \; . 
\end{equation}
Note that the local gross occupancies are the diagonal entries
of this matrix, $n_{\bbox{\gamma}}=C_{\bbox{\gamma},\bbox{\gamma}}$.
We express this matrix in the new basis,
\begin{equation}
C_{\bbox{\gamma_1},\bbox{\gamma'_1}} =
\sum_{\bbox{\sigma_1},\bbox{\sigma'_1}}
F_{\bbox{\gamma_1},\bbox{\sigma_1}}^{\vphantom{+}}
F_{\bbox{\sigma'_1},\bbox{\gamma'_1}}^+
\langle \hat{h}_{\bbox{\sigma_1}}^{+}
\hat{h}_{\bbox{\sigma'_1}}^{\vphantom{+}}\rangle \; . 
\end{equation}\end{mathletters}%
In infinite dimensions the entries of the interacting local
one-particle density matrix $\tensor{H}$ in the new basis
are readily calculated,
\begin{eqnarray}
H_{\bbox{\sigma_1},\bbox{\sigma'_1}} &=&
\langle \hat{h}_{\bbox{\sigma_1}}^{+}
\hat{h}_{\bbox{\sigma'_1}}^{\vphantom{+}}\rangle 
= 
\langle \hat{P}_{\text{G}} 
\hat{h}_{\bbox{\sigma_1}}^{+}
\hat{h}_{\bbox{\sigma'_1}}^{\vphantom{+}}
\hat{P}_{\text{G}} \rangle_0  \nonumber \\[3pt]
&=& 
\sum_{{\cal H}_1,{\cal H}_2,{\cal H}_3,{\cal H}_4}
\lambda_{{\cal H}_1,{\cal H}_2}
\lambda_{{\cal H}_3,{\cal H}_4}
\langle \hat{m}_{{\cal H}_1,{\cal H}_2}
\hat{h}_{\bbox{\sigma_1}}^{+}
\hat{h}_{\bbox{\sigma'_1}}^{\vphantom{+}}
\hat{m}_{{\cal H}_3,{\cal H}_4}
\rangle_0  \; .
\end{eqnarray}
In infinite dimensions we may set ${\cal H}_1={\cal H}_4={\cal H}'$,
${\cal H}_3={\cal H}\cup \bbox{\sigma'_1}$, and
${\cal H}_2={\cal H}\cup \bbox{\sigma_1}$ with
$\bbox{\sigma_1}, \bbox{\sigma'_1} \not\in {\cal H}$.
We then find 
\begin{eqnarray}
H_{\bbox{\sigma_1},\bbox{\sigma'_1}} &=&
\sum_{ 
{{\cal H}} \atop 
{{(\bbox{\sigma_1},\bbox{\sigma'_1} \not\in {\cal H})}}} \!\!
\text{fsgn}(\bbox{\sigma_1},{\cal H})
\text{fsgn}(\bbox{\sigma'_1},{\cal H})
\sum_{{\cal H}'} \lambda_{{\cal H}',{\cal H}\cup \bbox{\sigma_1}}
\lambda_{{\cal H}\cup \bbox{\sigma'_1},{\cal H}'} m_{{\cal H}'}^{h,0}
\nonumber \\[3pt]
&=&
\sum_{{ {\cal H} } \atop {(\bbox{\sigma_1},\bbox{\sigma'_1} \not\in {\cal H})}}
\!\! \text{fsgn}(\bbox{\sigma_1},{\cal H})
\text{fsgn}(\bbox{\sigma'_1},{\cal H})
\sum_{\Gamma,\Gamma'} 
\sqrt{\frac{m_{\Gamma}m_{\Gamma'}}{m_{\Gamma}^0m_{\Gamma'}^0}} 
A_{\Gamma,{\cal H}\cup\bbox{\sigma_1}}^+
A_{{\cal H}\cup\bbox{\sigma'_1},\Gamma'}^{\vphantom{+}}
\sum_{{\cal H}'} 
A_{{\cal H}',\Gamma}^{\vphantom{+}}
A_{\Gamma',{\cal H}'}^+ 
m_{{\cal H}'}^{h,0}
\; . \nonumber \\[3pt]
&& \label{Hlocalint}
\end{eqnarray}
Therefore, the matrix $\tensor{H}$ is known in terms of the 
variational parameters $m_{\Gamma}$ and the properties
of the one-particle product state~$|\Phi_0\rangle$.
If the non-interacting local density matrix $\tensor{C}^0$ is diagonal,
i.e., eq.~(\ref{noFOCKcops}) is fulfilled, the matrices~$\tensor{C}$
and $\tensor{H}$ are identical, and eq.~(\ref{Hlocalint}) 
also leads to~(\ref{hh1}).

The matrix $\tensor{H}$ is Hermitian and can thus be diagonalized
with the help of the unitary matrix $\tensor{X}$,
\begin{equation}
\left( \tensor{X}\right)^{+}\tensor{H} \tensor{X}=
\text{diag}(\widetilde{n}_{\bbox{\sigma}}^{h})\;,  
\label{matrixdefX}
\end{equation}
where $\widetilde{n}_{\bbox{\sigma}}^{h}$ are the eigenvalues
of the matrix~$\tensor{H}$. The entries of $\tensor{H}$ thus obey
\begin{eqnarray}
H_{\bbox{\sigma_1},\bbox{\sigma'_1}} &=&
\sum_{\bbox{\sigma_2},\bbox{\sigma'_2}}
X_{\bbox{\sigma_1},\bbox{\sigma_2}}^{\vphantom{+}}
\delta_{\bbox{\sigma_2},\bbox{\sigma'_2}}
\widetilde{n}_{\bbox{\sigma_2}}^{h} 
X_{\bbox{\sigma'_2},\bbox{\sigma'_1}}^{+}
\nonumber \\[3pt]
&=&
\sum_{\bbox{\sigma_2},\bbox{\sigma'_2}}
X_{\bbox{\sigma_1},\bbox{\sigma_2}}^{\vphantom{+}}
\frac{\widetilde{n}_{\bbox{\sigma_2}}^{h} }{n_{\bbox{\sigma_2}}^{h,0} }
\langle \hat{h}_{\bbox{\sigma_2}}^{+}
\hat{h}_{\bbox{\sigma'_2}}^{\vphantom{+}}\rangle_0
X_{\bbox{\sigma'_2},\bbox{\sigma'_1}}^{+}
\; . 
\end{eqnarray}
We transform back into the representation with $\hat{c}$~operators and
find for the interacting local one-particle density matrix in the 
original basis
\begin{equation}
C_{\bbox{\gamma_1},\bbox{\gamma'_1}}
=\sum_{\bbox{\gamma_2},\bbox{\gamma'_2}}
C_{\bbox{\gamma_2},\bbox{\gamma'_2}}^0
\sum_{\bbox{\sigma_1},\bbox{\sigma'_1},\bbox{\sigma_2},\bbox{\sigma'_2}}
F_{\bbox{\sigma_2},\bbox{\gamma_2}}^+
F_{\bbox{\gamma'_2},\bbox{\sigma'_2}}^{\vphantom{+}}
X_{\bbox{\sigma_1},\bbox{\sigma_2}}^{\vphantom{+}}
\frac{\widetilde{n}_{\bbox{\sigma_2}}^{h} }{n_{\bbox{\sigma_2}}^{h,0} }
X_{\bbox{\sigma'_2},\bbox{\sigma'_1}}^{+}
F_{\bbox{\gamma_1},\bbox{\sigma_1}}^{\vphantom{+}}
F_{\bbox{\sigma'_1},\bbox{\gamma'_1}}^+
\; . 
\label{intCmatrix}
\end{equation}
Recall that the local gross occupancies are the diagonal entries
of this matrix, $n_{\bbox{\gamma}}=C_{\bbox{\gamma},\bbox{\gamma}}$.

The result~(\ref{intCmatrix}) allows us to cast
the local hybridization term in the variational ground-state energy 
into the form
\begin{mathletters}
\begin{equation}
\sum_{\bbox{\gamma_1},\bbox{\gamma'_1}}
t^{\bbox{\gamma_1},\bbox{\gamma'_1}}
\langle \hat{c}_{\bbox{\gamma_1}}^{+}
\hat{c}_{\bbox{\gamma'_1}}^{\vphantom{+}}
\rangle =
\sum_{\bbox{\gamma_2},\bbox{\gamma'_2}}
\widetilde{t}^{\bbox{\gamma_2},\bbox{\gamma'_2}}
\langle \hat{c}_{\bbox{\gamma_2}}^{+}
\hat{c}_{\bbox{\gamma'_2}}^{\vphantom{+}}
\rangle_0 \; ,
\end{equation}
where the effective local hybridizations are given by
\begin{equation}
\widetilde{t}^{\bbox{\gamma_2},\bbox{\gamma'_2}}
= 
\sum_{\bbox{\gamma_1},\bbox{\gamma'_1}}
t^{\bbox{\gamma_1},\bbox{\gamma'_1}}
\sum_{\bbox{\sigma_1},\bbox{\sigma'_1}}
F_{\bbox{\gamma_1},\bbox{\sigma_1}}^{\vphantom{+}}
F_{\bbox{\sigma'_1},\bbox{\gamma'_1}}^+
\sum_{\bbox{\sigma_2},\bbox{\sigma'_2}}
X_{\bbox{\sigma_1},\bbox{\sigma_2}}^{\vphantom{+}}
\frac{\widetilde{n}_{\bbox{\sigma_2}}^{h} }{n_{\bbox{\sigma_2}}^{h,0} }
X_{\bbox{\sigma'_2},\bbox{\sigma'_1}}^{+}
F_{\bbox{\sigma_2},\bbox{\gamma_2}}^+
F_{\bbox{\gamma'_2},\bbox{\sigma'_2}}^{\vphantom{+}}
\; .
\end{equation}\end{mathletters}%

This expression simplifies if we assume that
there are no local Fock terms already in the
basis of the $\hat{c}$~operators. Then, the $\tensor{F}$~matrix becomes the
unit matrix. Let us further demand that orbitals with different 
crystal-field energies do not mix, i.e., $t^{\bbox{\gamma},\bbox{\gamma'}}
=\delta_{\bbox{\gamma},\bbox{\gamma'}}\epsilon_{\bbox{\gamma}}$.
If our one-particle product state~$|\Phi_0\rangle$ respects this
symmetry the $\tensor{X}$~matrix becomes the unit matrix, and we find
\begin{equation}
\widetilde{t}^{\bbox{\gamma_2},\bbox{\gamma'_2}}
= \delta_{\bbox{\gamma_2},\bbox{\gamma'_2}} \epsilon_{\bbox{\gamma_2}} 
\frac{n_{\bbox{\gamma_2}}}{n_{\bbox{\gamma_2}}^0} \; .
\end{equation}
Thus, we recover~(\ref{epsilontildedef}) for the effective crystal-field 
energies.

\begin{figure}[tbp]
\caption{Model density of states at the Fermi energy as a function of the
band filling $n_{\bbox{\sigma}}=n/4$. The dashed lines indicate the fillings
used in Sect.~\ref{twodegbands}. The total bandwidth is~$W=6.6$ eV.}
\label{DOSfig}
\end{figure}

\begin{figure}[tbp]
\caption{Variational parameters as a function of~$U$ for $J=0.2 U$ 
($U'=0.6 U$) at half band-filling.}
\label{varparameter}
\end{figure}

\begin{figure}[tbp]
\caption{Bandwidth renormalization factor~$q$ at half band-filling
as a function of~$U$ for various values of~$J$ ($J=(U-U')/2$).}
\label{qfuncUJ}
\end{figure}

\begin{figure}[tbp]
\caption{Bandwidth renormalization factor at the
Brinkman--Rice transition $U=U_{\text{BR}}$ as a function of~$U'/U$ 
for the Gutzwiller wave function with atomic correlations
(full line) and pure density correlations (dotted line).
Also shown is the value of the $q$~factor for GW$_{\text{atom}}$ at 
$U=U_{\text{BR}}^{\text{dens}}$ (dashed line).}
\label{qdiscontinuity}
\end{figure}

\begin{figure}[tbp]
\caption{Phase diagram and critical interaction strength for the
Brinkman--Rice transition in Gutzwiller wave functions with atomic
(GW$_{\text{atom}}$) and pure density (GW$_{\text{dens}}$) 
correlations as a function of
$U'/U$ (left Y-axis). 
The dashed curve shows the energy gain for atomic correlations against pure
density correlations at $U=U_{\text{BR}}^{\text{dens}}$ (right Y-axis).}
\label{phasedia}
\end{figure}

\begin{figure}[tbp]
\caption{Size of the local spin 
$\langle \bigl(\hat{\vec{S}}_i\bigr)^2\rangle $ 
in the paramagnetic Gutzwiller wave function with atomic
correlations as a function of the interaction strength and various values of 
$J=(U-U')/2$ for half band-filling.}
\label{Ssquared}
\end{figure}

\begin{figure}[tbp]
\caption{Magnetization density per band as a function of~$U$ for $J=0.2U$
for the Hartree--Fock solution (HF), the Gutzwiller wave function with
pure density correlations (GW$_{\text{dens}}$), and the Gutzwiller wave function
with atomic correlations (GW$_{\text{atom}}$) 
for (a)~$n/4=0.29$ and (b)~$n/4=0.35$.
The dotted line indicates 
the results for GW$_{\text{atom}}$ with $J_{\text{C}}=0$.}
\label{magnetizationdensity}
\end{figure}

\begin{figure}[tbp]
\caption{Size of the local spin $\langle \bigl(\hat{\vec{S}}_i\bigr)^2
\rangle $ as a function of the interaction strength for $J=0.2U$ and
band-filling $n/4=0.35$ for the Hartree--Fock theory (HF) and the Gutzwiller
wave functions (GW$_{\text{dens}}$, GW$_{\text{atom}}$).}
\label{Localspinferro}
\end{figure}

\begin{figure}[tbp]
\caption{Phase diagram as a function of~$U$ and~$J$ for the Hartree--Fock
solution (HF) and the two Gutzwiller wave functions
(GW$_{\text{dens}}$, GW$_{\text{atom}}$) 
for (a)~$n/4=0.29$ and (b)~$n/4=0.35$; PM: paramagnet, FM: ferromagnet.}
\label{Phasediaferro}
\end{figure}

\begin{figure}[tbp]
\caption{Condensation energy as a function of~$U$ for $J=0.2U$ for the
Hartree--Fock theory (HF) and the Gutzwiller wave function 
(GW$_{\text{atom}}$) for $n_{\sigma}=n/4=0.29$ (full lines) 
and~$n_{\sigma}=n/4=0.35$ (dashed lines).}
\label{condenenergy}
\end{figure}

\mediumtext

\begin{table}[tbp]
\caption{Eigenstates with symmetry specifications, spin
quantum numbers, energies, and notation symbols
for the sixteen $N=2$ atomic configurations}
\label{tableone}
\begin{tabular}{ccccccc}
\# & Atomic eigenstate $|\Gamma \rangle $ & Symmetry & $S_{\text{at}}$ & 
$S_{\text{at}}^z$ & energy $E_{\Gamma} $ & prob. \\ 
\tableline 1 & $|0,0\rangle $ & $a_1$ & 0 & 0 & 0 & $e$ \\ 
2 & $|\uparrow ,0\rangle $ & $e_g$ & 1/2 & 1/2 & 0 & $s_{\uparrow }$ \\ 
3 & $|0,\uparrow \rangle $ & $e_g$ & 1/2 & 1/2 & 0 & $s_{\uparrow }$ \\ 
4 & $|\downarrow ,0\rangle $ & $e_g$ & 1/2 & $-$1/2\hphantom{$-$} & 0 & 
$s_{\downarrow }$ \\ 
5 & $|0,\downarrow \rangle $ & $e_g$ & 1/2 & $-$1/2\hphantom{$-$} & 0 & 
$s_{\downarrow }$ \\ 
6 & $|\uparrow ,\uparrow \rangle $ & ${}^3A_2$ & 1 & 1 & $U'-J$ &
$d_t^{\uparrow \uparrow }$ \\ 
7 & $(|\uparrow ,\downarrow \rangle +|\downarrow ,\uparrow \rangle )/\sqrt{2}
$ & ${}^3A_2$ & 1 & 0 & $U'-J$ & $d_t^{\text{0}}$ \\ 
8 & $|\downarrow ,\downarrow \rangle $ & ${}^3A_2$ & 1 & $-$1\hphantom{$-$}
& $U'-J$ & $d_t^{\downarrow \downarrow }$ \\ 
9 & $(|\uparrow ,\downarrow \rangle -|\downarrow ,\uparrow \rangle )/\sqrt{2}
$ & ${}^1E$ & 0 & 0 & $U'+J$ & $d_E$ \\ 
10 & $(|\uparrow \downarrow ,0\rangle -|0,\uparrow \downarrow \rangle )/\sqrt{2}$
 & ${}^1E$ & 0 & 0 & $U-J_{\text{C}}$ & $d_E$ \\ 
11 & $(|\uparrow \downarrow ,0\rangle +|0,\uparrow \downarrow \rangle )/\sqrt{2}$
 & ${}^1A_1$ & 0 & 0 & $U+J_{\text{C}}$ & $d_A$ \\ 
12 & $|\uparrow ,\uparrow \downarrow \rangle $ & $E_g$ & 1/2 & 1/2 & 
$U+2U'-J$ & $t_{\uparrow }$ \\ 
13 & $|\uparrow \downarrow ,\uparrow \rangle $ & $E_g$ & 1/2 & 1/2 & 
$U+2U'-J$ & $t_{\uparrow }$ \\ 
14 & $|\downarrow ,\uparrow \downarrow \rangle $ & $E_g$ & 1/2 & 
$-$1/2\hphantom{$-$} & $U+2U'-J$ & $t_{\downarrow }$ \\ 
15 & $|\uparrow \downarrow ,\downarrow \rangle $ & $E_g$ & 1/2 & 
$-$1/2\hphantom{$-$} & $U+2U'-J$ & $t_{\downarrow }$ \\ 
16 & $|\uparrow \downarrow ,\uparrow \downarrow \rangle $ & $A_1$ & 0 & 0 & 
$2U+4U'-2J$ & $f$
\end{tabular}
\end{table}

\begin{table}[tbp]
\caption{Two-electron spin-orbit states with spin quantum numbers, energies,
and notation symbols for the case of pure density correlations}
\label{tabletwo}
\begin{tabular}{ccccc}
\# & Wave function $|I\rangle $ & $S_{\text{at}}^z$ & energy $U_I$ & prob.
\\ 
\tableline 6 & $|\uparrow ,\uparrow \rangle $ & 1 & $U'-J$ & 
$d_{1}^{\uparrow\uparrow}$ \\ 
7 & $|\downarrow ,\downarrow \rangle $ & $-$1\hphantom{$-$} & $U'-J$
& $d_{1}^{\downarrow,\downarrow }$ \\ 
8 & $|\downarrow ,\uparrow \rangle $ & 0 & $U'$ & $d_s$ \\ 
9 & $|\uparrow ,\downarrow \rangle $ & 0 & $U'$ & $d_s$ \\ 
10 & $|\uparrow \downarrow ,0\rangle $ & 0 & $U$ & $d_c$ \\ 
11 & $|0,\uparrow \downarrow \rangle $ & 0 & $U$ & $d_c$
\end{tabular}
\end{table}


\begin{references}
\bibitem{info2}  In Ref.~\onlinecite{BGW} we used the notation~$\eta_{i;I}$
instead of~$g_{i;I}$ for $|I|\leq 1$.
In Ref.~\onlinecite{GebhardPRB1} the notation $g^{-\mu_{i;\uparrow }}
\equiv g_{i;\uparrow }$, $g^{-\mu_{i;\downarrow }}\equiv
g_{i;\uparrow }$, and $g^{\eta_i}\equiv g_{i;\emptyset }$ was used.
\end{references}
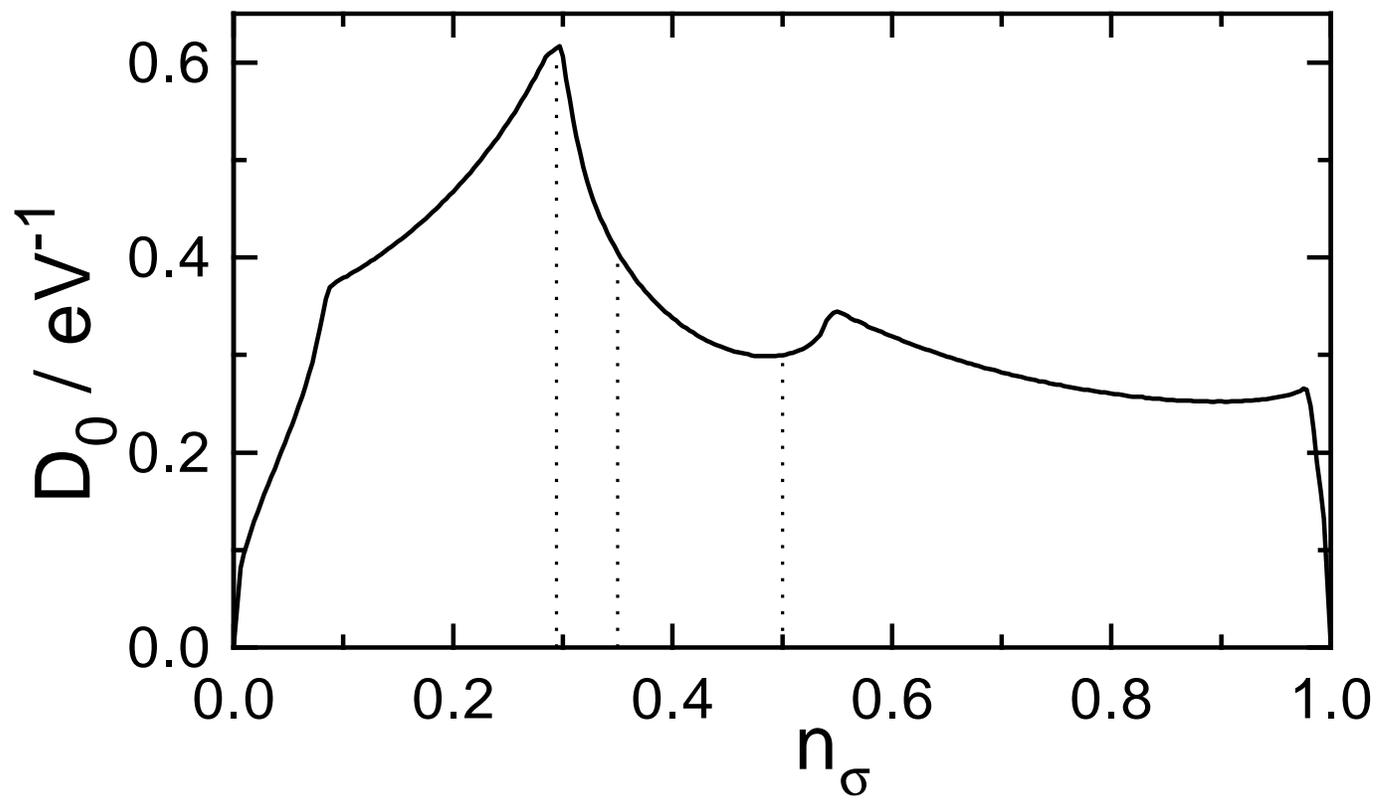
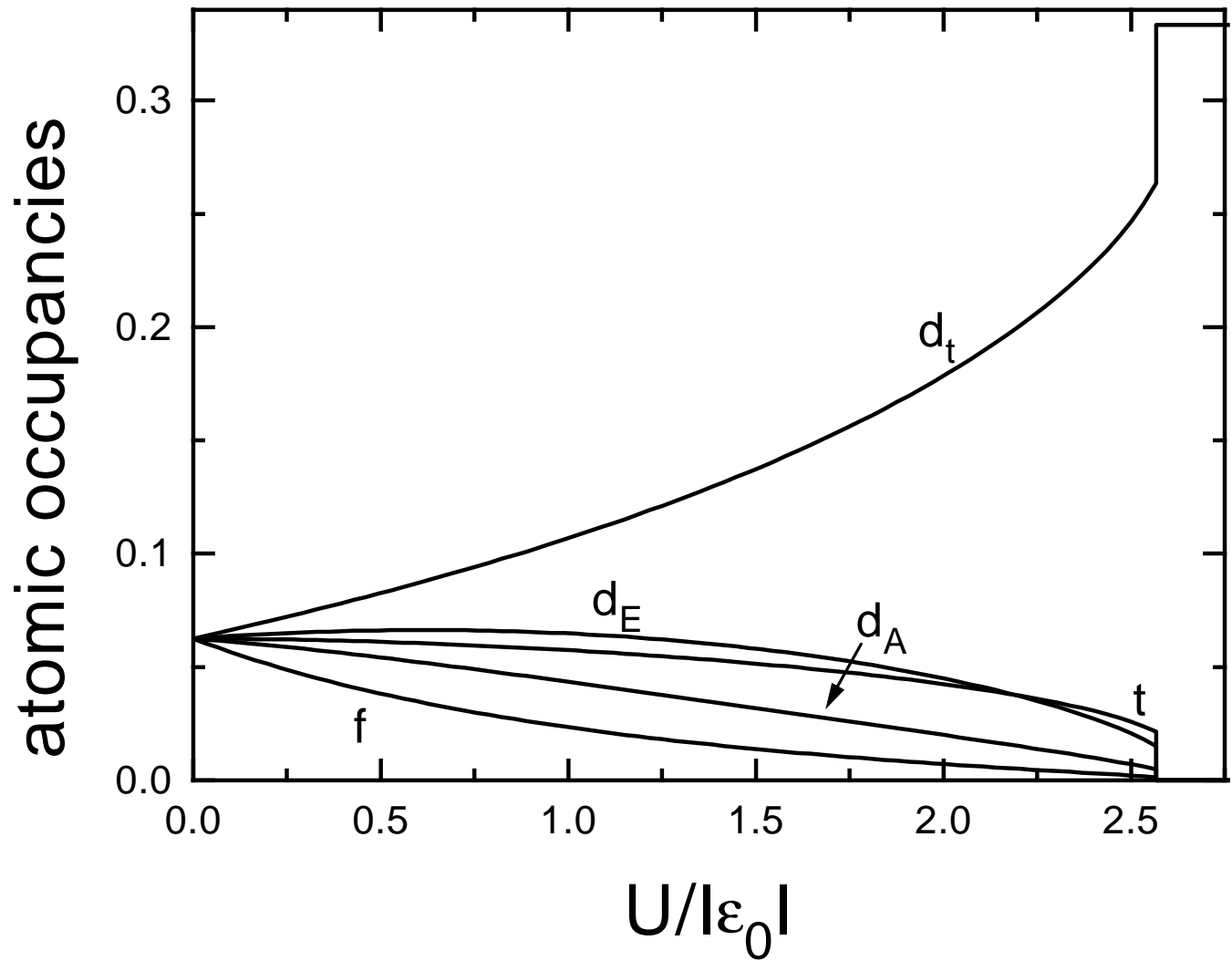
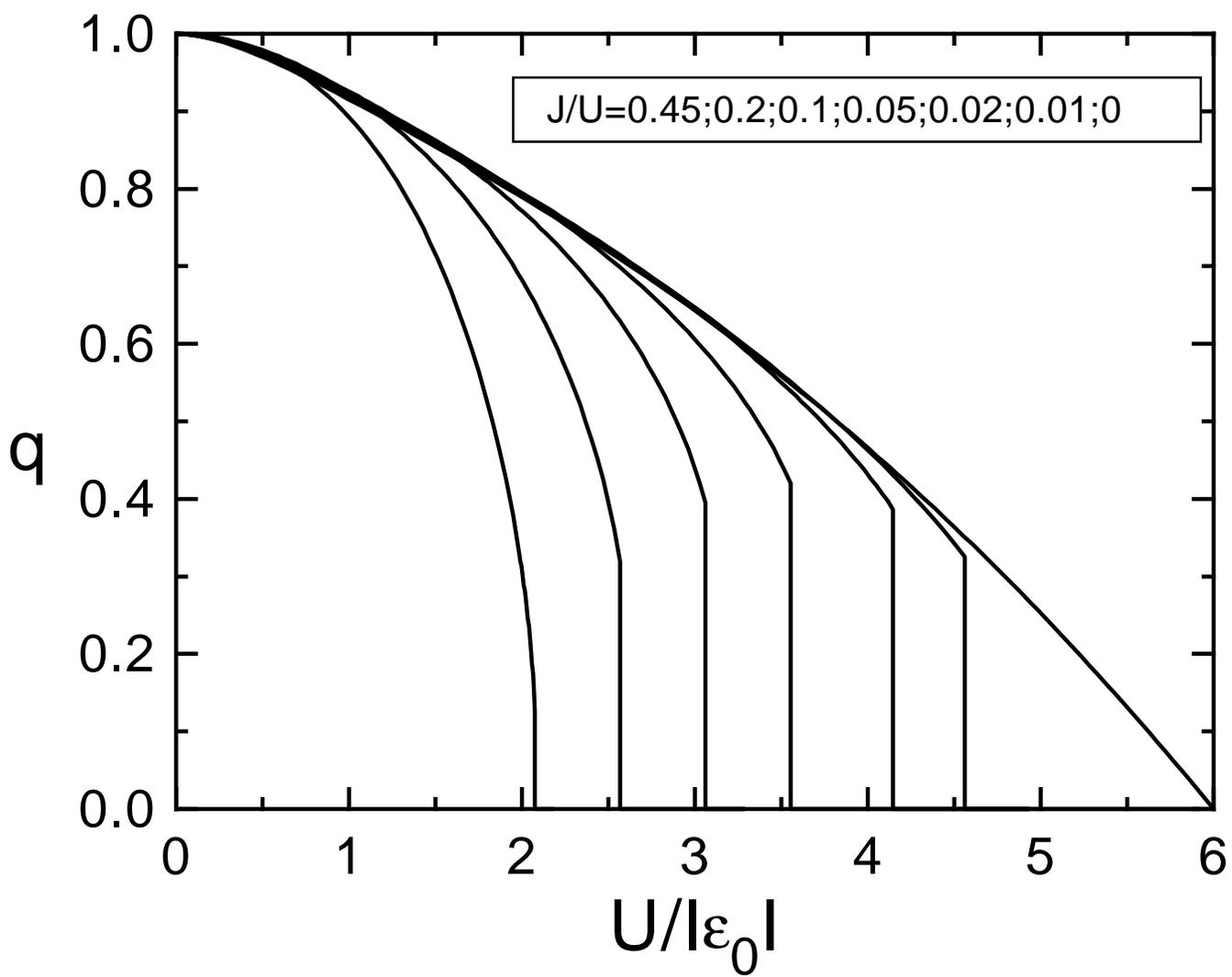
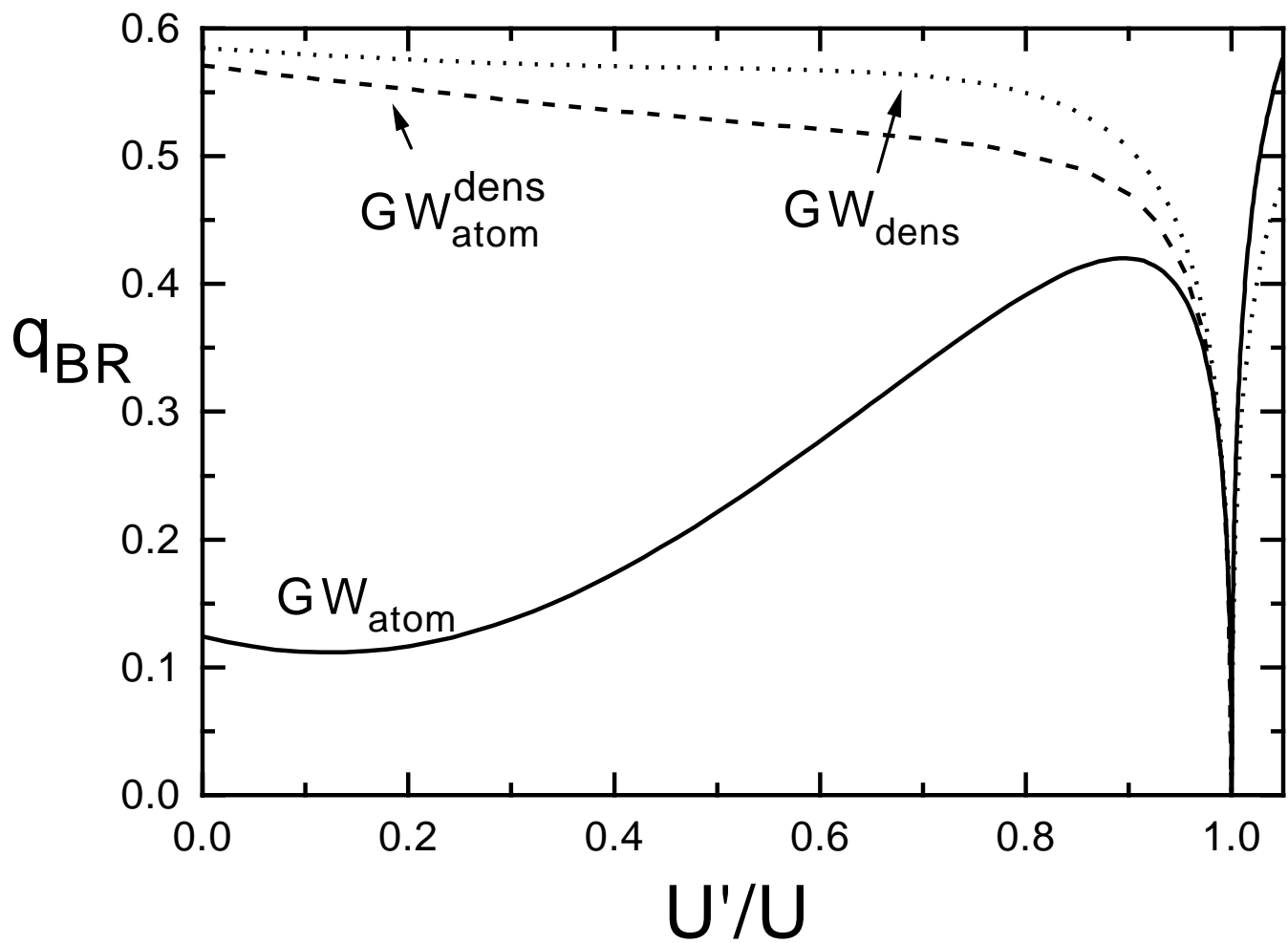
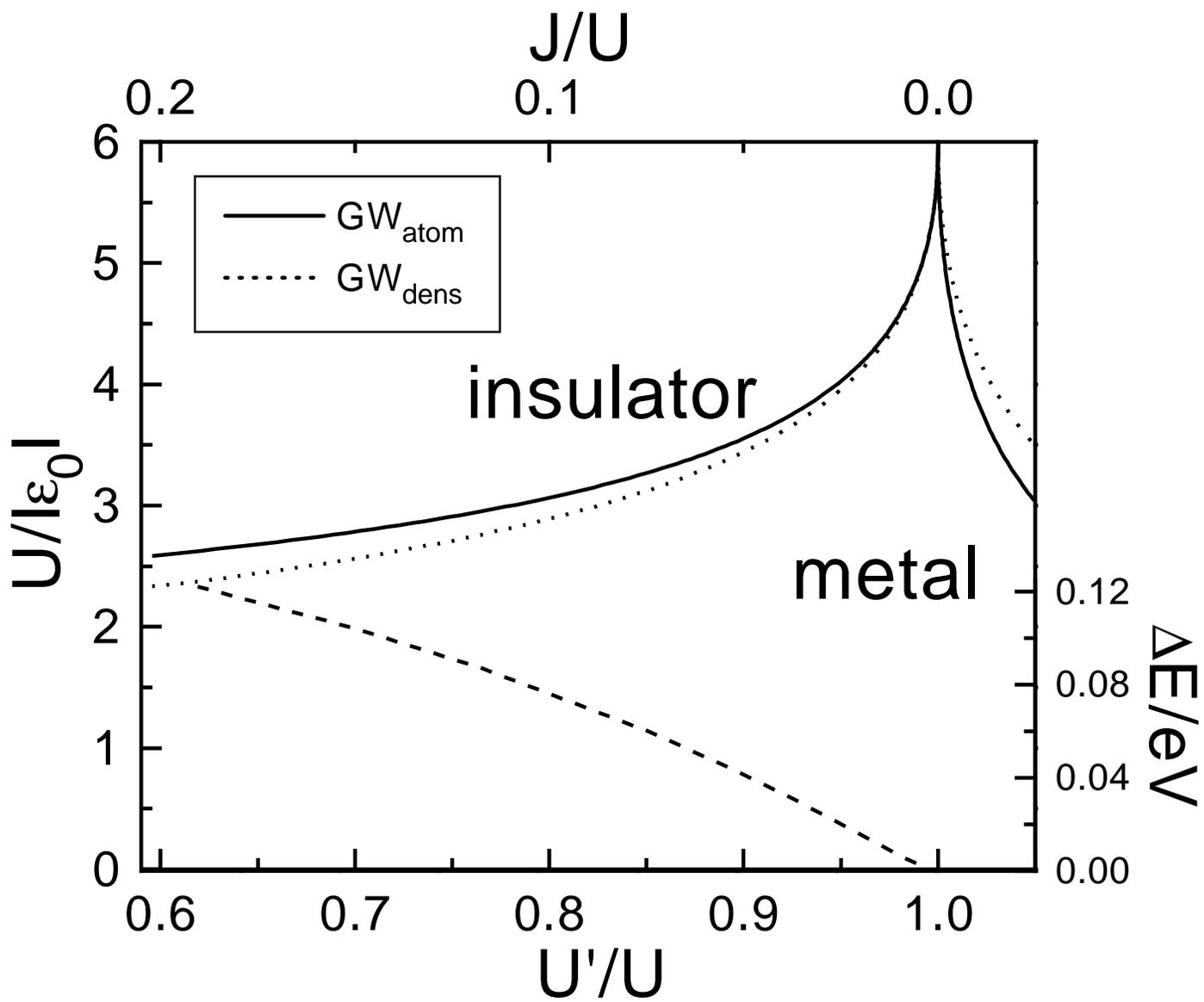
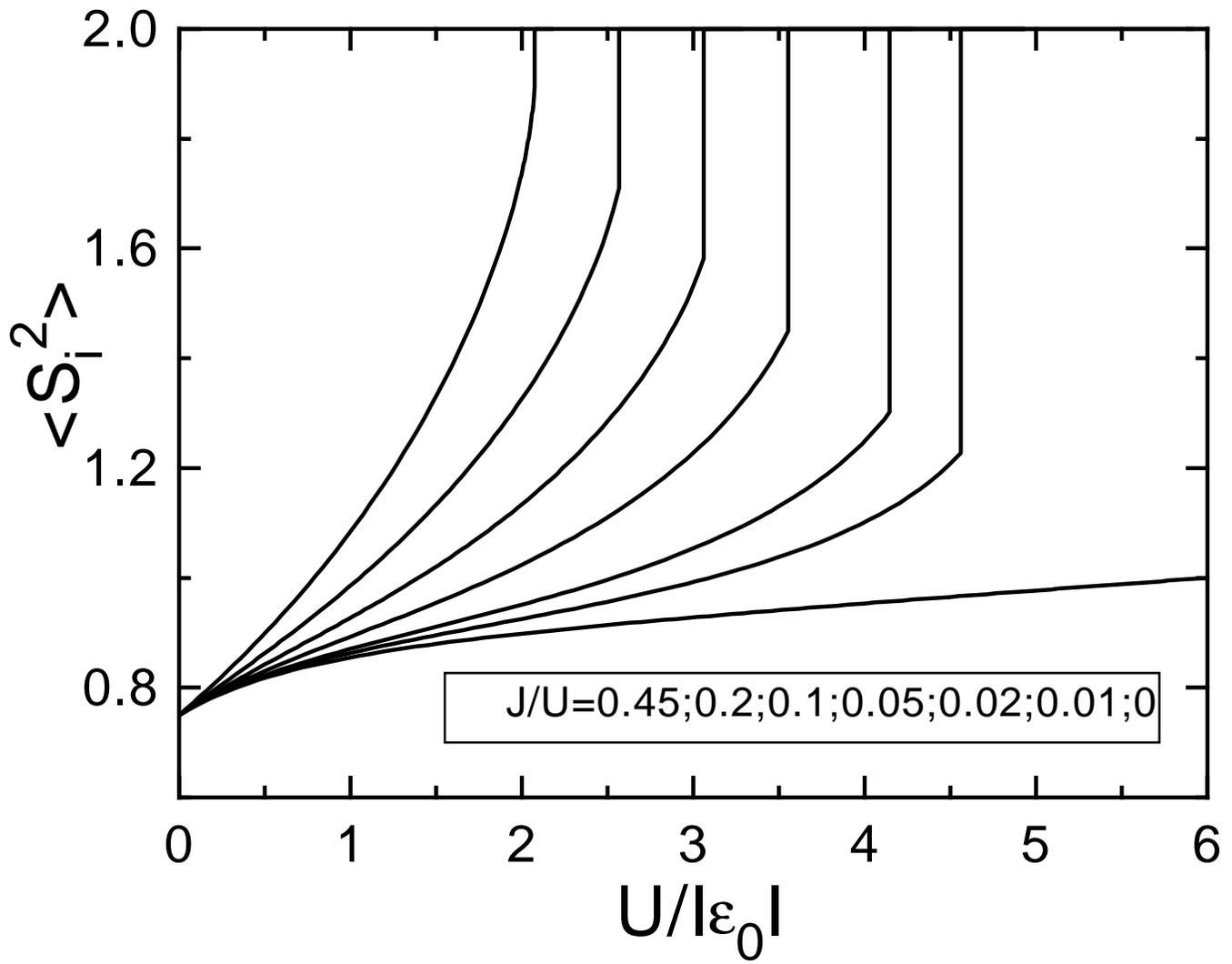
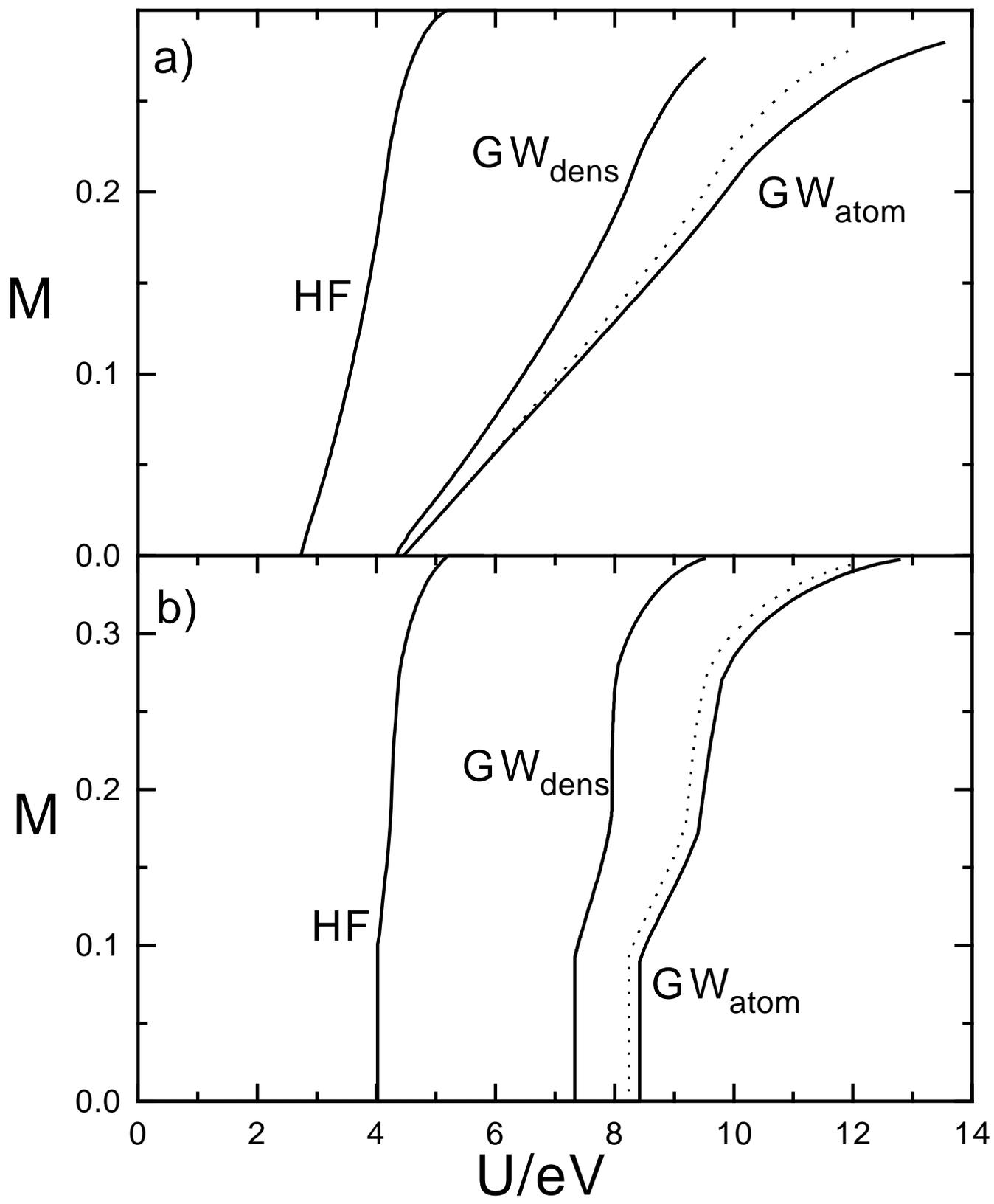
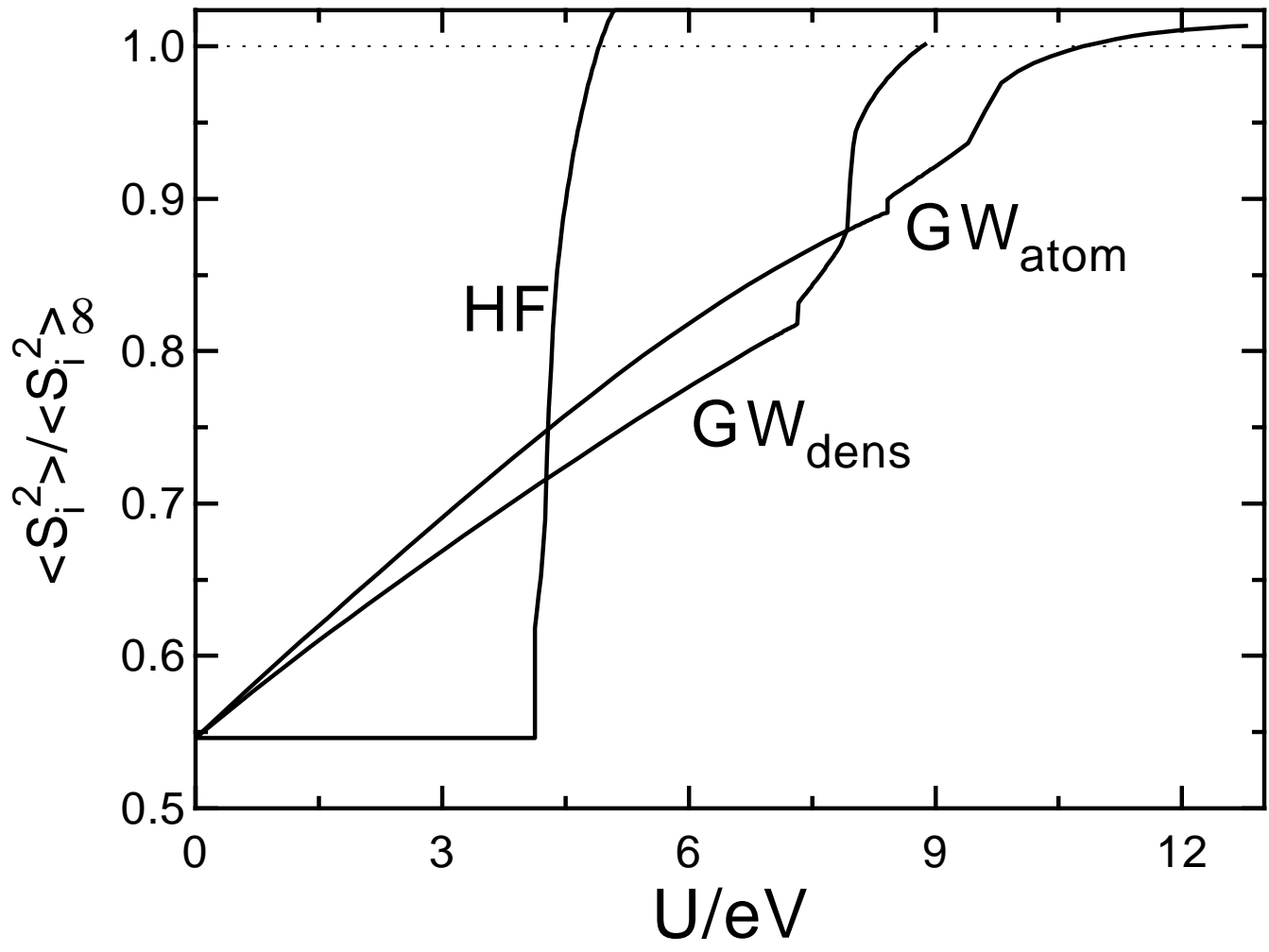
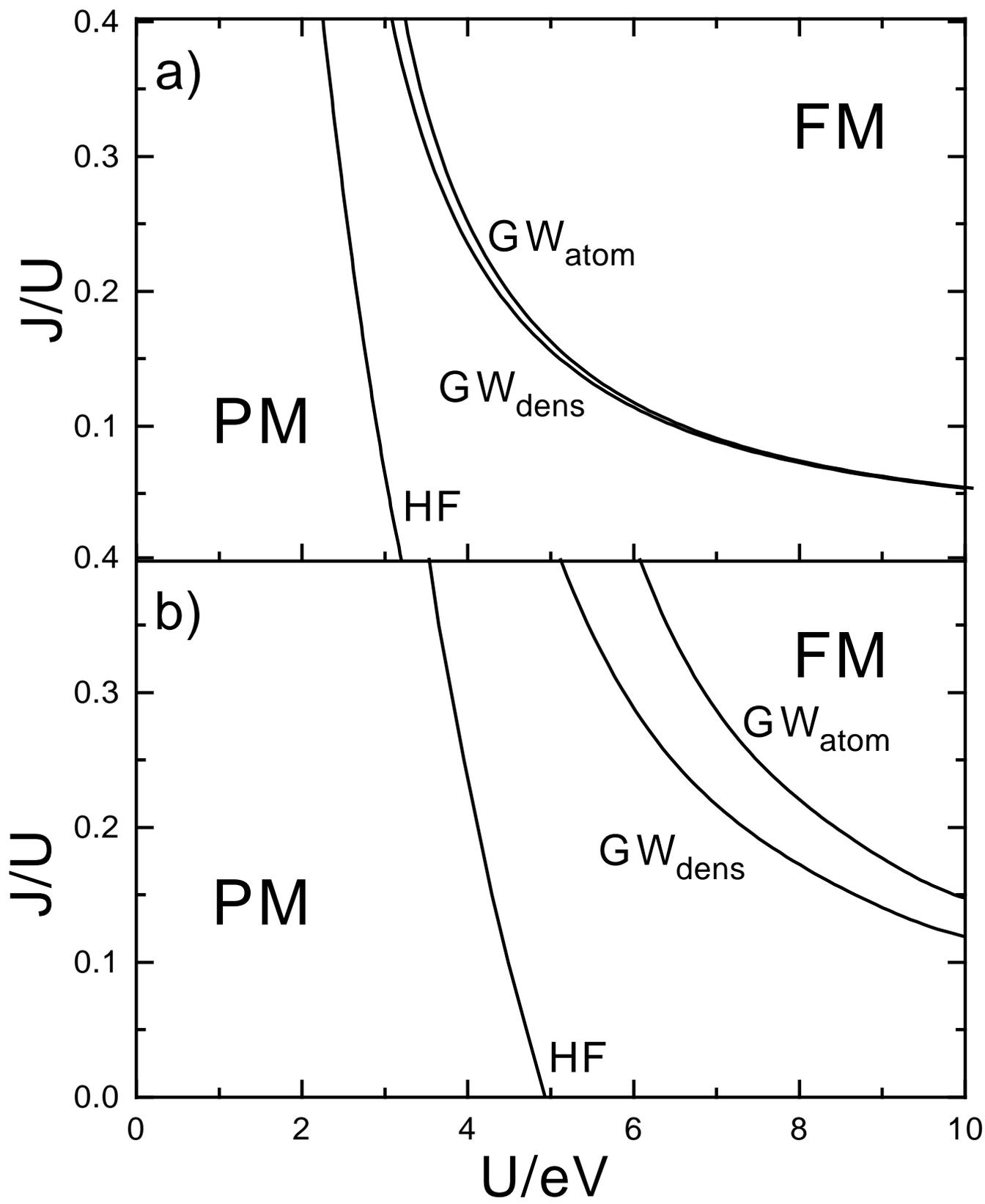
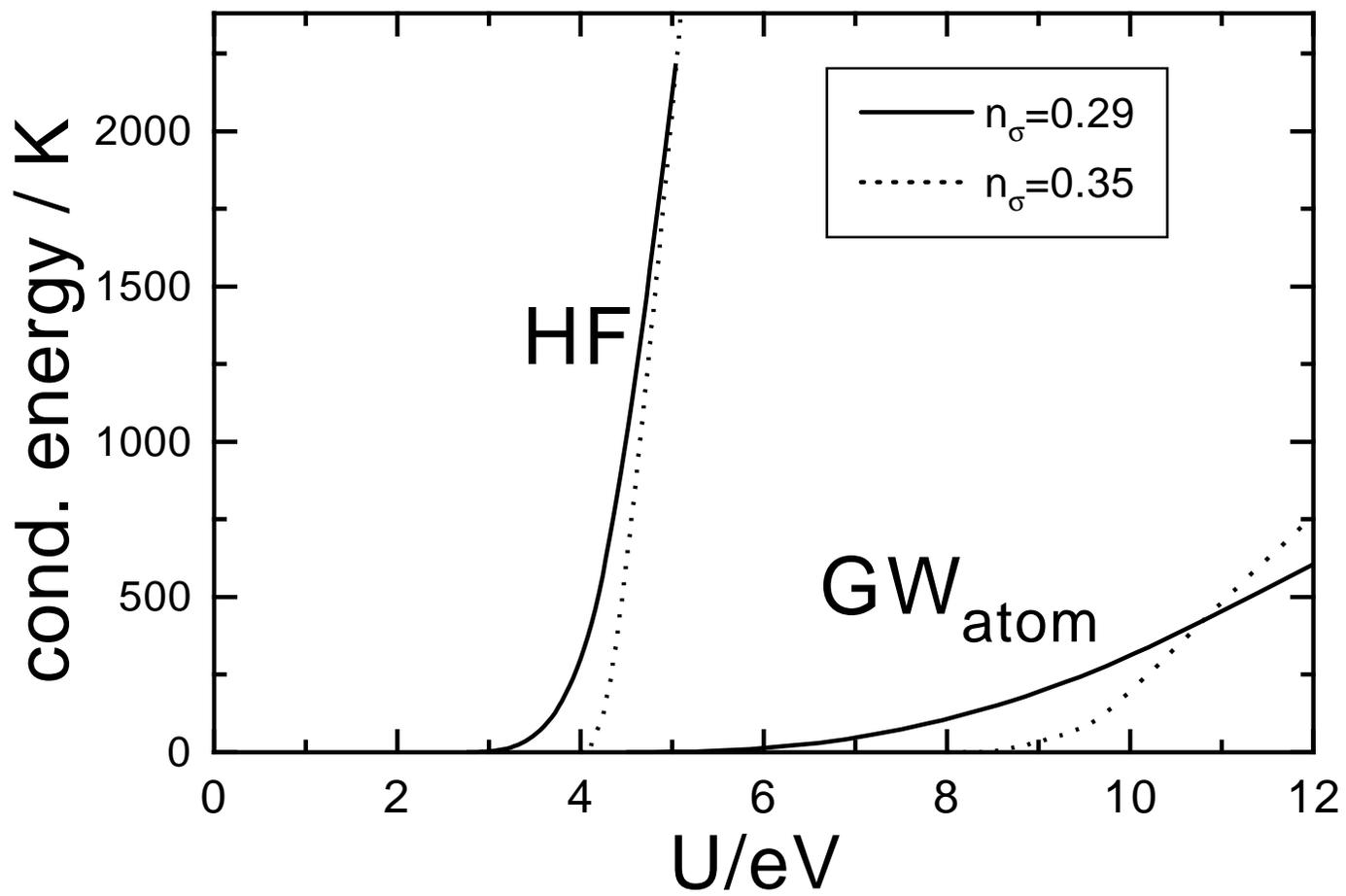
\end{document}